\documentclass[12pt]{amsart}
\usepackage{amsfonts}
\usepackage{latexsym,amsmath,amsthm,amssymb,amsfonts}

\numberwithin{equation}{section}
\allowdisplaybreaks

\def\endproof{$\hfill\Box$\\}
\def\s{\,\,\,\,}

\numberwithin{equation}{section}
\newtheorem{theorem}{Theorem}[section]
\newtheorem{lem}[theorem]{Lemma}
\newtheorem{thm}[theorem]{Theorem}
\newtheorem{pro}[theorem]{Proposition}

\newtheorem{rem}[theorem]{Remark}

\newcounter{Cnumber}

\title[ ]
{\bf Global existence of the solution to Einstein-Yang-Mills-Higgs equations with small initial datum}
\author[ ]
{Zonglin Jia \qquad Boling Guo }

\date{}
\begin{document}
\maketitle

\begin{abstract}
The problem involved in this paper is the global existence of the solution to the $\mathfrak{su}(2)$-Einstein-Yang-Mills-Higgs(EYMH) equation. The approach we employ stems from H. Lindblad and I. Rodnianski and is dependent of wave coordinates and Lorentzian gauge conditions. Our main conclusion is that the EYMH system admits global existence provided the initial datum are sufficiently small. To the best of our knowledge, there is no similar result in the area of EYMH equations.

\end{abstract}

\section{Introduction}
Recently many mathematicians are concerned about the EYMH equation. They, following the idea of D. Christodoulou (see e.g. \cite{C1986}), usually reduce this equation under special metrics to get global existence. In our present article, we take into account of general metric solving the EYMH system and are going to get global existence with small initial datum.

This paper is inspired by the work in \cite{LR2010}, which give a new proof of the global stability of Minkowski space originally established by Christodoulou and Klainerman in \cite{CK1993}. The smart method of H. Lindblad and I. Rodnianski is based on the wave coordinates, which play a critical role in giving some more exquisite estimates. Following their idea, we also employ wave coordinates and Lorentzian gauges. Like \cite{LR2010} our frame of the article is the contradiction argument.

Throughout the paper, the same indices appearing twice means summing it. Besides, we also appoint that, when denoting superscripts or  subscripts, the Greek letters such as $\alpha,\beta,\gamma,\cdots$ belong to $\{0,1,2,3\}$, while the Latin letters $i,j,k,\cdots$ are in $\{1,2,3\}$.

We consider the following equations on $(\mathbb{R}^4,g)$
\begin{eqnarray*}
\left\{
\begin{array}{rcl}
&Ric_{\alpha\beta}-\frac{1}{2}R\cdot g_{\alpha\beta}=T_{\alpha\beta}        & \quad\quad\text{(Einstein equation)}\\
&g^{\lambda\mu}(\nabla_{\lambda}\tilde{F}_{\mu\alpha}+[A_{\lambda},\tilde{F}_{\mu\alpha}])=\tilde{J}_{\alpha}      & \quad\quad\text{(Yang-Mills equation)}\\
&g^{\lambda\mu}\hat{\nabla}_{\lambda}\hat{\nabla}_{\mu}\Phi=V'(|\Phi|^2)\Phi,     & \quad\quad\text{(Higgs equation)}
\end{array} \right.
\end{eqnarray*}
where $g$ is a Lorentzian metric with signature $(-,+,+,+)$, $Ric$ and $R$ are Ricci tensor and scalar curvature of $g$ respectively. Furthermore, we assume that $\tilde{F}_{\alpha\beta}:=\partial_{\alpha}A_{\beta}-\partial_{\beta}A_{\alpha}+[A_{\alpha},A_{\beta}]$ and $T_{\alpha\beta}$ is given by
\begin{eqnarray}\label{9}
T_{\alpha\beta}&:=&\langle \tilde{F}_{\alpha}^{\lambda},\tilde{F}_{\beta\lambda}\rangle-\frac{1}{4}g_{\alpha\beta}\langle \tilde{F}^{\lambda\mu},\tilde{F}_{\lambda\mu}\rangle+(\hat{\nabla}_{\alpha}\Phi)^{\dag}\hat{\nabla}_{\beta}\Phi+(\hat{\nabla}_{\beta}\Phi)^{\dag}\hat{\nabla}_{\alpha}\Phi\\
&&-g_{\alpha\beta}\{g^{\rho\sigma}(\hat{\nabla}_{\rho}\Phi)^{\dag}\hat{\nabla}_{\sigma}\Phi+V(\Phi^{\dag}\Phi)\},\nonumber
\end{eqnarray}
where we have raised the indices by $(g^{\alpha\beta})$ which is the inverse of the metric matrix $(g_{\alpha\beta})$. $A$ is a 1-form, called the Yang-Mills potential, defined on $\mathbb{R}^4$ with values in the Lie algebra $\mathfrak{su}(2)$; $\Phi\in C^1(\mathbb{R}^4,\mathbb{C}^2)$ is a scalar-multiplet field, called the Higgs field. $\nabla$ denotes the covariant derivative with respect to the metric $g$, and $[\cdot,\cdot]$ means the Lie bracket of the Lie algebra $\mathfrak{su}(2)$. Moreover, $\hat{\nabla}_{\alpha}\Phi$ is the gauge covariant derivative of the complex doublet Higgs field, defined by
\begin{eqnarray*}
\hat{\nabla}_{\alpha}\Phi:=\partial_{\alpha}\Phi-\sqrt{-1}A^I_{\alpha}\sigma_I\Phi/2,
\end{eqnarray*}
where $(\sigma_I)_{I=1,2,3}$ are the conventional Pauli spin matrices. $\Phi^{\dag}$ denotes the Hermitian conjugate of $\Phi$ and $|\Phi|$ is given by $|\Phi|^2:=\Phi^{\dag}\Phi$. $V$ is a real function defined on $\mathbb{R}^+$, often called the self-interaction potential, with the $s$-th derivative $V^{(s)}$. It is worthy to point out that $\mathfrak{su}(2)$ admits a non-degenerate inner product, denoted by the symbol $``\langle\cdot,\cdot\rangle"$, satisfying the following property
\begin{eqnarray}\label{97}
\langle f,[k,l]\rangle=\langle[f,k],l\rangle\s\s\s\s\forall f,k,l\in\mathfrak{su}(2).
\end{eqnarray}
$\tilde{J}$ is a $\mathfrak{su}(2)$-valued one-form called the Yang-Mills current, defined by
\begin{eqnarray*}
\tilde{J}:=\{\Phi^{\dag}S^I\hat{\nabla}_{\alpha}\Phi-(\hat{\nabla}_{\alpha}\Phi)^{\dag}S^I\Phi\}dx^{\alpha}\otimes T_I,
\end{eqnarray*}
where $S^I:=\sqrt{-1}\sigma_I/2$ and $\{T_I\}_{I=1,2,3}$ is the basis of $\mathfrak{su}(2)$ given by
\begin{eqnarray*}
T_1:=\sqrt{-1}\sigma_1/2,\s\s\s\s T_2:=-\sqrt{-1}\sigma_2/2\s\s\s\s\mbox{and}\s\s\s\s T_3:=\sqrt{-1}\sigma_3/2.
\end{eqnarray*}

In order to make our main result easier to understand, we have to do some preparation firstly. Let $h:=g-m$, where $m$ is the Minkowski metric of $\mathbb{R}^4$, and set
\begin{eqnarray*}
\mathfrak{L}:=\{\partial_{\alpha},\s\Omega_{\alpha\beta}:=-x_{\alpha}\partial_{\beta}+x_{\beta}\partial_{\alpha},\s \tilde{S}:=t\partial_t+r\partial_r\}.
\end{eqnarray*}
This series of vector fields plays a vital role in the research of wave equations in Minkowski space-time. We denote the above vector fields by $Z^{\iota}$, where $\iota:=(0,\cdots,1,\cdots,0)$. Let $I:=(\iota_1,\cdots,\iota_k)$ with $|\iota_i|=1$ for $1\leqslant i\leqslant k$, be a multi-index of length $|I|=k$ and let $Z^I:=Z^{\iota_1}\cdots Z^{\iota_k}$ denote a product of $k$ vector fields from the family $\mathfrak{L}$. By a sum $I_1+I_2=I$ we mean a sum over all possible order preserving partitions of the multi-index $I$ into two multi-indices $I_1$ and $I_2$, i.e. if $I=(\iota_1,\cdots,\iota_k)$, then $I_1=(\iota_{i_1},\cdots,\iota_{i_n})$ and $I_2=(\iota_{i_{n+1}},\cdots,\iota_{i_k})$, where $i_1,\cdots,i_k$ is any reordering of the integers $1,\cdots,k$ such that $i_1<\cdots<i_n$ and $i_{n+1}<\cdots<i_k$. The usual wave operator is given by $\Box:=m^{\alpha\beta}\partial_{\alpha}\partial_{\beta}$. From Section 2 of \cite{LR2010} it follows that we have the next commutation properties:
\begin{eqnarray*}
[\Box,\partial_{\alpha}]=0,\s\s\s\s[\Box,\Omega_{\alpha\beta}]=0\s\s\s\s\mbox{and}\s\s\s\s[\Box,\tilde{S}]=2\Box,
\end{eqnarray*}
where $[X,Y]:=XY-YX$ is the commutator. For $Z\in\mathfrak{L}$, denote $[Z,\Box]:=c_Z\cdot\Box$, i.e. $c_Z=2$, if $Z=\tilde{S}$, and 0, otherwise. Moreover, we set
\begin{equation}\label{95}
\mathcal{E}_N(t):=\sum\limits_{|I|\leqslant N,Z\in\mathfrak{L}}\left(||\sqrt{w}\partial Z^Ih(\cdot,t)||_{L^2}+||\sqrt{w}\partial Z^IA(\cdot,t)||_{L^2}+||\sqrt{w}\partial Z^I\Phi(\cdot,t)||_{L^2}\right),
\end{equation}
where the weight is defined as
\begin{eqnarray*}
w(q):=
\left\{
\begin{array}{rcl}
&1+(1+|q|)^{1+2\gamma}        & \quad\quad\text{when $q>0$,}\\
&1+(1+|q|)^{-2\mu}         & \quad\quad\text{when $q<0$}
\end{array} \right.
\end{eqnarray*}
with $q:=|x|-t$ and two constants $\gamma\in(0,1/2)$ and $\mu\in(0,1/2)$ being fixed.\\

Now we state the main result of this paper.
\begin{thm}\label{thm6.1}
Given an integer $N\geqslant4$, there exists a constant $\varepsilon_0>0$ such that if $\varepsilon\leqslant\varepsilon_0$ and the initial datum $h|_{t=0}$, $\partial_th|_{t=0}$, $A|_{t=0}$, $\partial_tA|_{t=0}$, $\Phi|_{t=0}$ and $\partial_t\Phi|_{t=0}$ obey $\mathcal{E}_N(0)\leqslant\varepsilon$, then the solution of EYMH equations $(g(t)=h(t)+m,A(t),\Phi(t))$ belongs to $\tilde{E}_{N+1}(\infty)$, provided $|V^{(s)}(x)|\leqslant C_s\cdot x^{\tilde{\gamma}}$ for any integer $s\in[0,N-4]$ and any $x\in[0,1]$. Here $\tilde{\gamma}\in(3/2+\gamma,\infty)$ is a fixed constant.
\end{thm}

\begin{rem}
The definition of $\tilde{E}_{N+1}(\infty)$ is given in Subsection \ref{subsection2.6}.
\end{rem}

In the sequel, we would like to review some previous results. In \cite{C2001}, D. Chae considered the characteristic initial-value problem of the coupled Einstein and nonlinear Klein-Gordon system, where data is given on an initial outgoing null hypersurface, and under spherically symmetric assumption he studied the global evolution problem toward future of the given initial null hypersurface. Employing similar tricks, in \cite{C2003} D. Chae gave a detailed proof of global existence of solutions to Einstein-Maxwell-Higgs system in the spherical symmetry. In \cite{TT2012}, C. Tadmon and S.B. Tchapnda investigated the initial value problem for the spherically symmetric $\mathfrak{su}(2)$-EYMH system. They obtained global existence and decay properties. In addition, people care about relativistic hydrodynamics. In \cite{ST2014}, M. Sango and C. Tadmon considered the Einstein-Maxwell-Euler and obtained global well-posedness in Bondi coordinates. All the idea of \cite{C2001,C2003,ST2014,TT2012} origins from \cite{C1986} and is to reduce the system to a single first order integro-differential equation and then use the contraction mapping theorem in appropriate function spaces.

For another Einstein-matter system there is a semi-global result for the Einstein-Maxwell-Yang-Mills equations for small data due to Friedrich \cite{F1991}. His results are based on analysis of the ``conformal structure" of the Einstein field equations, i.e. on a study of the ``conformal Einstein equations" which must be satisfied by the ``nonphysical" metric which is obtained from the ``physical" metric by a conformal rescaling. The ``semi-global" is in the sense that all its ``physical" null geodesies are past complete. In \cite{D2003}, M. Dafermos studied the stability and instability of the Cauchy horizon for the spherically symmetric Einstein-Maxwell-scalar field equations and resolved the issue of uniqueness in the context of a special, spherically symmetric initial value problem for a system of gravity coupled with matter, whose relation to the problem of gravitational collapse is well established in the physics literature. His result is related to the strong cosmic censorship conjecture of Roger Penrose. In \cite{JG}, Z. Jia and B. Guo investigated the Einstein-Maxwell(EM) equations and got that this system admits a global solution with small initial datum. In \cite{SWY1993}, J.A. Smoller, A.G. Wasserman and S.T. Yau shew that the $\mathfrak{su}(2)$-Einstein-Yang-Mills(EYM) equations admit an infinite family of ``black-hole" solutions having a regular event horizon, for every choice of the radius of the event horizon. In \cite{SW1997}, J.A. Smoller and A.G. Wasserman proved the following property of spherically symmetric solutions to the $\mathfrak{su}(2)$-EYM equations: Any solution to the EYM system which is defined in the far field($r>>1$) and has finite ADM mass, is defined for all $r\in(0,\infty)$. This fact is surprising, since in general for nonlinear equations, existence theorems are usually only local. However, for these equations they got a global existence for all solutions defined in a neighborhood of infinity.

In the case of non-relativistic, there are also research results. In \cite{Y2015}, S. Yang studied the asymptotic behavior of solutions to the Maxwell-Klein-Gordon(MKG) equations on $\mathbb{R}^4$ with large Cauchy data. In order to get strong quantitative decay estimates for solutions, he brought in some weighted energy space. In \cite{YY2018}, S. Yang and P. Yu provided a gauge independent proof of a conjecture, which states that the solutions to MKG equations disperse as linear waves and enjoy peeling properties for pointwise estimates. The remarkable works \cite{EM1} and \cite{EM2} of Eardley and Moncrief established the global existence to the general Yang-Mills-Higgs system with sufficiently smooth initial datum. The key step is the demonstration that the $L^{\infty}$-norm of the curvature is bounded a priori. Their results apply to any compact gauge group and to any invariant Higgs self-coupling which is positive and of no higher than quartic degree.

Now let us briefly introduce the method we use. Firstly, in the process of getting local well-posedness, we applied wave coordinates and Lorentzian gauges to transform the EYMH equations into a hyperbolic system called the reduced EYMH systems. In order to show that the solution to the reduced EYMH systems also solves the original EYMH equations, we have to require that the initial datum sets satisfy EYMH constraints. For the details of the above concepts readers may refer to Section \ref{section2} of this paper. Secondly, as soon as we get a local solution, it is natural to consider the maximal existence time $T$ and assume it to be finite. In the next, we define $T^*$ to be
\begin{eqnarray}\label{00}
T^*:=\sup\{T_0:\exists C=C(T),\,s.t.\s\forall t\in[0,T_0),\s\mathcal{E}_N(t)\leqslant2C\varepsilon\}
\end{eqnarray}
and suppose that $T^*<T$. We will show that if $\varepsilon>0$ is small enough, then the inequality in (\ref{00}) implies the same inequality with $2C$ replaced by $C$ for all $t<T^*$. This contradicts the maximality of $T^*$ and we will obtain that the inequality holds for all $t<T$. Moreover, since the energy $\mathcal{E}_N(t)$ is now finite at $t=T$(Note that $\mathcal{E}_N(t)$ is continuous with respect to $t$), we can extend the solution beyond $T$ to contradict maximality of $T$ and show that $T=\infty$. Hence, our final task is to get energy estimates. Noting the definition of $\mathcal{E}_N$, we compute $\overset{\sim}{\Box}_gZ^Ih$, $\overset{\sim}{\Box}_gZ^IA$ and $\overset{\sim}{\Box}_gZ^I\Phi$ where $\overset{\sim}{\Box}_g:=g^{\alpha\beta}\partial_{\alpha}\partial_{\beta}$. Applying Proposition 6.2 of \cite{LR2010} and Gronwall inequality leads to the needed results.

This paper is organized as follows:

We devote Section \ref{section2} to some preliminaries. In Section \ref{section3} we reduce the EYMH equation to a hyperbolic system under wave coordinates and Lorentzian gauge conditions and get local existence and uniqueness. In Section \ref{section4} we rewrite the reduced EYMH equation as another one with respect to $(h:=g-m,A,\Phi)$. Moreover, decay estimates and energy estimates are given in Section \ref{section5} and Section \ref{section6} respectively.

\section{Notations and Preliminaries}\label{section2}
In this article, the symbol ``$Q_1\lesssim Q_2$" means that there exists a constant $C$ such that $Q_1\leqslant C\cdot Q_2$ for two given quantities $Q_1$ and $Q_2$. Throughout our paper, the constant $C$ may depend upon the maximal existence time $T$.
\subsection{Equivalence of the Einstein equation}
It is easy to check that the Einstein equation is equivalent to a simpler one
\begin{equation}\label{0}
Ric_{\alpha\beta}=\langle \tilde{F}_{\alpha}^{\lambda},\tilde{F}_{\beta\lambda}\rangle-\frac{1}{4}g_{\alpha\beta}\langle \tilde{F}^{\lambda\mu},\tilde{F}_{\lambda\mu}\rangle+(\hat{\nabla}_{\alpha}\Phi)^{\dag}\hat{\nabla}_{\beta}\Phi+(\hat{\nabla}_{\beta}\Phi)^{\dag}\hat{\nabla}_{\alpha}\Phi+g_{\alpha\beta}V(|\Phi|^2).
\end{equation}
Thanks to the above transformation, we can rewrite a complicated equation as a hyperbolic system in wave coordinates later.
\subsection{Wave coordinates}
We say that a metric $g$ of $\mathbb{R}^4$ satisfies the wave coordinates condition if
\begin{eqnarray*}
\hat{F}^{\lambda}:=g^{\alpha\beta}\Gamma^{\lambda}_{\alpha\beta}=0,
\end{eqnarray*}
where $\Gamma^{\lambda}_{\alpha\beta}$ is the connection coefficient of $g$.
\subsection{Uniform equivalence of Riemannian metric}
Given two Riemannian metric $e_1$ and $e_2$ on a smooth manifold $M$, we say they are uniformly equivalent to each other if there exist two constants $0<A<B<\infty$ such that for all $X\in TM$, the following holds true
\begin{eqnarray*}
A\cdot e_1(X,X)\leqslant e_2(X,X)\leqslant B\cdot e_1(X,X).
\end{eqnarray*}
\subsection{Sliced spacetime}
A Lorentzian spacetime $(V_T:=\mathbb{R}^3\times[0,T),g)$ is called sliced for some $t\in(0,\infty]$ if there exists a time-dependent vector $\beta(x,t):=\beta^i(x,t)\frac{\partial}{\partial x^i}(x)$, which is called the shift, tangent to the space slice $M_t:=\mathbb{R}^3\times\{t\}$ such that

(1) $\vec{n}(\cdot,t):=\frac{\partial}{\partial t}(\cdot,t)-\beta(\cdot,t)$ is a normal vector to $M_t$ for all $t\in[0,T)$. That is to say, given $X_t\in T(M_t)$($T(M_t)$ is the tangent bundle of $M_t$), we have
\begin{eqnarray}\label{5}
g(\vec{n}(\cdot,t),X_t)=0.
\end{eqnarray}
It is easy to check that (\ref{5}) is equivalent to $g_{0i}=g_{ij}\beta^j$;

(2) $\vec{n}$ is timelike, namely,
\begin{eqnarray}\label{6}
g(\vec{n},\vec{n})<0.
\end{eqnarray}
Combining (\ref{5}) and (\ref{6}) we arrive at $g_{ij}\beta^i\beta^j>g_{00}$.

Using the above inequality we define a positive function $N$ called the lapse which is given by $N:=\sqrt{g_{ij}\beta^i\beta^j-g_{00}}$. Then we can write $g$ as
\begin{eqnarray}\label{11}
g=-N^2dt\otimes dt+g_{ij}(dx^i+\beta^idt)\otimes(dx^j+\beta^jdt).
\end{eqnarray}
Because $g$ is Lorentzian, $(g_{ij})$ is positive definite. In order to represent $(g^{\alpha\beta})$ via $(N,\beta,g_{ij})$, we denote the inverse of $(g_{ij})$ by $(g_{*}^{ij})$. It is easy to check
\begin{eqnarray*}
g^{00}=-N^{-2},\s\s\s\s g^{0i}=N^{-2}\beta^i\s\s\s\s\mbox{and}\s\s\s\s g^{ij}=g_*^{ij}-N^{-2}\beta^i\beta^j.
\end{eqnarray*}

Thanks to the above discussion, now we can give the following lemma without proof.
\begin{lem}\label{lem2.3}
A spacetime $(V_T,g)$ is sliced and Lorentzian if and only if $g_t:=i_t^*g$, which is induced by the embedding $i_t:\mathbb{R}^3\longrightarrow V_T$, $x\mapsto(x,t)$, is positive definite and
\begin{eqnarray*}
g_{0i}\cdot g_{*}^{ij}\cdot g_{0j}>g_{00}.
\end{eqnarray*}
\end{lem}

\begin{rem}
Lemma \ref{lem2.3} tells us that determining a sliced Lorentzian metric $g$ on $V_T$ is equivalent to determining the following quantities:

(1) a Riemannian metric $g_t$ on $M_t$;

(2) a positive function $N$ on $M_t$;

(3) a tangent vector field $\beta$ to $M_t$.
\end{rem}

\begin{rem}
In case $g_t$ is positive definite, the inverse of $(g_t)_{ij}$ which is denoted by $(g_t)^{ij}$ is just $g_*^{ij}$. Hence, throughout this article we always use the symbol $g_*^{ij}$.
\end{rem}

\begin{rem}
Let ${}^t\Gamma^k_{ij}$ be the coefficient of the Levi-Civita connection ${}^t\nabla$ on $(M_t,g_t)$. Now we are going to give the specific expression of $\Gamma^{\theta}_{\lambda\mu}$ via ${}^t\Gamma^k_{ij}$, $\beta^i$ and $N$. They are
\begin{eqnarray*}
\Gamma^k_{ij}=\frac{1}{2}N^{-2}\beta^k(g_{jp}\cdot{}^t\nabla_i\beta^p+g_{ip}\cdot{}^t\nabla_j\beta^p-\partial_tg_{ij})+{}^t\Gamma^k_{ij};
\end{eqnarray*}
\begin{eqnarray*}
\Gamma^0_{ij}=\frac{1}{2}N^{-2}\partial_tg_{ij}-\frac{1}{2}N^{-2}(g_{kj}\cdot{}^t\nabla_i\beta^k+g_{ki}\cdot{}^t\nabla_j\beta^k);
\end{eqnarray*}
\begin{eqnarray*}
\Gamma^k_{0j}=\frac{1}{2}(g_*^{kl}-N^{-2}\beta^k\beta^l)\cdot(\partial_tg_{lj}+{}^t\nabla_j\beta^qg_{ql}-{}^t\nabla_l\beta^qg_{qj})+\frac{\beta^k}{2N^2}\partial_j\{g(\beta,\beta)-N^2\};
\end{eqnarray*}
\begin{eqnarray*}
\Gamma^k_{00}=\frac{1}{2}\left(g_*^{kl}-\frac{\beta^k\beta^l}{N^2}\right)\{2\partial_t(g_{lp}\beta^p)-\partial_l[g(\beta,\beta)-N^2]\}+\frac{\beta^k}{2N^2}\partial_t\{g(\beta,\beta)-N^2\};
\end{eqnarray*}
\begin{eqnarray*}
\Gamma^0_{i0}=N^{-1}\partial_iN-\frac{1}{4}N^{-2}\partial_i\{g(\beta,\beta)\}-\frac{1}{2}N^{-2}\beta^l\cdot{}^t\nabla_l\beta^pg_{pi}+\frac{1}{2}N^{-2}\beta^l\partial_tg_{il};
\end{eqnarray*}
\begin{eqnarray*}
\Gamma^0_{00}=\frac{\beta^l\beta^p\partial_tg_{pl}}{2N^2}+\frac{(\partial_t+\beta)(N)}{N}-\frac{(L_{\beta}g)(\beta,\beta)}{2N^2},
\end{eqnarray*}
where $L_{\beta}$ is the Lie derivative with respect to $\beta$.
\end{rem}

\subsection{Regular sliced spacetime}
A sliced spacetime $(V_T,g,\beta)$ is called regular with respect to $(\mathbb{R}^3,e)$(where $e$ is the standard Euclidean metric of $\mathbb{R}^3$) if

(1) The metrics $g_t$ are uniformly equivalent to $e$, i.e. there exist continuous strictly positive functions $B_1(t)$, $B_2(t)$ such that for all $t\in[0,T)$ and each tangent vector $X$ to $\mathbb{R}^3$ it holds true on $M_t$
\begin{eqnarray*}
B_1(t)\cdot e(X,X)\leqslant g_t(X,X)\leqslant B_2(t)\cdot e(X,X);
\end{eqnarray*}

(2) The lapse $N$ is such that there exist continuous strictly positive functions $C_1(t)$, $C_2(t)$ on $[0,T)$ such that on each $M_t$ it holds true
\begin{eqnarray*}
C_1(t)\leqslant N(x,t)\leqslant C_2(t)\s\s\s\s\forall x\in\mathbb{R}^3;
\end{eqnarray*}

(3) The shift $\beta$ is uniformly bounded in $e$-norm on each $M_t$ by a number $b(t)$.
\subsection{Sobolev space on $V_T$}\label{subsection2.6}
We denote by $E_s(T)$ the following Banach space
$$E_s(T):=\bigcap\limits_{0\leqslant k\leqslant s}C^{s-k}([0,T),H_k(\mathbb{R}^3)).$$

The Sobolev space $\tilde{E}_s(T)$ is the space of functions $u$, such that $u\in\bar{C}(V_T)$, space of continuous and bounded functions on $V_T$, while $\partial u\in E_{s-1}(V_T)$, where $\partial$ is the Levi-Civita connection on $(V_T,m)$(Recall that $m$ is the standard Minkowski metric of $V_T\subseteq\mathbb{R}^4$).
\subsection{EYMH initial data set}\label{sub2.8}
A EYMH initial data set is a ten-tuple\\$(\mathbb{R}^3;\bar{g},K,\bar{A}^{\mbox{space}},\bar{A}_0, \bar{E}, \bar{\beta},\bar{N},\bar{\Phi},\bar{\Psi})$. $\bar{g}$ and $K$ are the Riemannian metric and symmetric 2-tensor on $\mathbb{R}^3$ respectively. Meanwhile, $\bar{A}^{\mbox{space}}$ is a $\mathfrak{su}(2)$-valued one-form on $\mathbb{R}^3$; $\bar{A}_0$ is a $\mathfrak{su}(2)$-valued function on $\mathbb{R}^3$; $\bar{E}$ is a $\mathfrak{su}(2)$-valued vector field on $\mathbb{R}^3$; $\bar{\beta}$ is a vector field on $\mathbb{R}^3$. Moreover, $\bar{N}$ is a positive function on $\mathbb{R}^3$. $\bar{\Phi}$ and $\bar{\Psi}$ are $\mathbb{C}^2$-valued functions on $\mathbb{R}^3$, where $\mathbb{C}$ is the set of all the complex numbers.
\subsection{EYMH development}
The development of the initial data set $(\mathbb{R}^3;\bar{g},K,\bar{A}^{\mbox{space}},\bar{\Phi})$ is a tetrad $(\mathbb{R}^4,g,A,\Phi)$ with a Lorentzian metric $g$, a $\mathfrak{su}(2)$-valued one-form $A$ on $\mathbb{R}^4$ and a $\mathbb{C}^2$-valued function $\Phi$ on $\mathbb{R}^4$, such that the embedding $i_0$ of $\mathbb{R}^3$ into $\mathbb{R}^4$(Recall that $i_0(x):=(x,0)$ for all $x\in\mathbb{R}^3$) enjoys the following properties:

(a) The metric $\bar{g}$ is the pullback of $g$ by $i_0$, i.e. $\bar{g}=i_0^*g$;

(b) $K$ is the second fundamental form of $i_0(\mathbb{R}^3)$ as a submanifold of $(\mathbb{R}^4,g)$;

(c) The $\mathfrak{su}(2)$-valued one-form $\bar{A}^{\mbox{space}}$ is the pullback of $A$ by $i_0$, i.e. $\bar{A}^{\mbox{space}}=i_0^*A$;

(d) $\bar{\Phi}(x)=\Phi(x,0)$ for all $x\in\mathbb{R}^3$.

Furthermore, $(g,A,\Phi)$ satisfies on $\mathbb{R}^4$ the EYMH equations and $(\mathbb{R}^4,g)$ is a sliced spacetime.
\subsection{EYMH constraints}
Restricting EYMH equations to the initial data set\\$(\mathbb{R}^3;\bar{g},K,\bar{A}^{\mbox{space}},\bar{A}_0, \bar{E}, \bar{\beta},\bar{N},\bar{\Phi})$, which is embedded into $\mathbb{R}^4$, leads to the following identities called the constraints
\begin{eqnarray}\label{1}
\bar{R}-|K|^2_{\bar{g}}+(\mbox{tr}_{\bar{g}}K)^2=2\bar{N}^{-2}\cdot T(\partial_t-\beta,\partial_t-\beta)|_{t=0},
\end{eqnarray}

\begin{eqnarray}\label{2}
(\mbox{div}_{\bar{g}}K)_i-\partial_i(\mbox{tr}_{\bar{g}}K)=-\bar{N}^{-1}\cdot T(\partial_t-\beta,\partial_i)|_{t=0},
\end{eqnarray}

\begin{eqnarray}\label{3}
&&\tilde{J}(\beta-\partial_t)|_{t=0}=\bar{g}^{ij}(\bar{\nabla}_i\bar{F}_{jk}+[\bar{A}^{\mbox{space}}_i,\bar{F}_{jk}])\bar{\beta}^k+\bar{N}\cdot[\bar{A}^{\mbox{space}}_i,\bar{E}^i]+\bar{F}(\bar{\beta},\bar{\nabla}_{\bar{\beta}}\bar{\beta})/\bar{N}^2\nonumber\\
&&-2\bar{\beta}(\bar{N})\bar{g}(\bar{\beta},\bar{E})/\bar{N}^2+\bar{E}(\bar{N})+\bar{N}\mbox{div}_{\bar{g}}\bar{E}+2\bar{g}(\bar{E},\bar{\beta})\mbox{tr}_{\bar{g}}K+\bar{g}^{ij}\bar{F}_{ik}\bar{\nabla}_j\bar{\beta}^k\\
&&+\bar{F}(\bar{\beta},\mbox{grad}_{\bar{g}}\bar{N})/\bar{N}-2\bar{F}_{ik}\bar{\beta}^i\bar{\beta}^jK_{jl}\bar{g}^{kl}/\bar{N},\nonumber
\end{eqnarray}

\begin{eqnarray}\label{10}
\partial_tA_i|_{t=0}-\partial_i\bar{A}_0+[\bar{A}_0,\bar{A}^{\mbox{space}}_i]:=\bar{N}\cdot\bar{E}^j\cdot\bar{g}_{ji}
\end{eqnarray}

\begin{eqnarray}\label{98}
(\partial_t\Phi)|_{t=0}-\sqrt{-1}(\bar{A}_0)^I\sigma_I\bar{\Phi}/2:=\bar{\Psi}
\end{eqnarray}
where $\bar{F}:=d\bar{A}^{\mbox{space}}+[\bar{A}^{\mbox{space}},\bar{A}^{\mbox{space}}]$, $\bar{\nabla}$ and $\bar{R}$ are the Levi-Civita connection and the scalar curvature of $(\mathbb{R}^3,\bar{g})$ respectively.
\begin{rem}
Unless we give (\ref{10}) and (\ref{98}), one can not determine the values on the right hand side of (\ref{1}), (\ref{2}) and (\ref{3}). In the next, we are going to write their specific expressions via $(\bar{g},\bar{N},\bar{F},\bar{E},\bar{\beta},\bar{\Phi},\bar{\Psi})$:
\begin{eqnarray*}
&&2\bar{N}^{-2}\cdot T(\partial_t-\beta,\partial_t-\beta)|_{t=0}=\bar{g}^{ik}\bar{g}^{jl}\langle\bar{F}_{ij},\bar{F}_{kl}\rangle/2+2\bar{g}_{ij}\langle\bar{E}^i,\bar{E}^j\rangle+2|\bar{\Psi}-\hat{\rlap{/}\nabla}_{\bar{\beta}}\bar{\Phi}|^2/\bar{N}^2\\
&&+\frac{\bar{g}^{ij}}{\bar{N}^2}\langle\bar{\beta}^k\bar{F}_{ki}-\bar{N}\bar{E}^l\bar{g}_{li},\bar{\beta}^k\bar{F}_{kj}-\bar{N}\bar{E}^l\bar{g}_{lj}\rangle+2\{\bar{g}^{ij}(\hat{\rlap{/}\nabla}_i\bar{\Phi})^{\dag}(\hat{\rlap{/}\nabla}_j\bar{\Phi})+V(|\bar{\Phi}|^2)\}
\end{eqnarray*}
and
\begin{eqnarray*}
-\bar{N}^{-1}\cdot T(\partial_t-\beta,\partial_i)|_{t=0}&=&\langle\bar{E}^l,\bar{F}_{li}\rangle+\bar{N}^{-1}\bar{\beta}^k\bar{g}^{pq}\langle\bar{F}_{pk},\bar{F}_{qi}\rangle-\bar{N}^{-1}(\bar{\Psi}-\bar{\beta}^k\hat{\rlap{/}\nabla}_k\bar{\Phi})^{\dag}\hat{\rlap{/}\nabla}_i\bar{\Phi}\\
&&-\bar{N}^{-1}(\hat{\rlap{/}\nabla}_i\bar{\Phi})^{\dag}(\bar{\Psi}-\bar{\beta}^k\hat{\rlap{/}\nabla}_k\bar{\Phi}),
\end{eqnarray*}
where $\hat{\rlap{/}\nabla}$ is given by
\begin{eqnarray*}
\hat{\rlap{/}\nabla}_X\bar{\Phi}:=X(\bar{\Phi})-\frac{\sqrt{-1}}{2}(\bar{A}^{\mbox{space}}(X))^I\sigma_I\bar{\Phi}
\end{eqnarray*}
and $X$ is a tangent vector field on $\mathbb{R}^3$. Moreover, we also have
\begin{eqnarray*}
\tilde{J}(\beta-\partial_t)|_{t=0}=\bar{\Phi}^{\dag}S(\bar{\beta}^k\hat{\rlap{/}\nabla}_k\bar{\Phi}-\bar{\Psi})-(\bar{\beta}^k\hat{\rlap{/}\nabla}_k\bar{\Phi}-\bar{\Psi})^{\dag}S\bar{\Phi},
\end{eqnarray*}
where for any $\mathbb{C}^2-$valued function $\psi$, we define $S\psi:=(S^I\psi)\cdot T_I$. Recall that $S^I:=\sqrt{-1}\sigma_I/2$ and $(T_I)_{I=1,2,3}$ is the basis of $\mathfrak{su}(2)$.
\end{rem}

\subsection{The null frame}\label{sub2.11}
At each point $x\in\mathbb{R}^4$, we introduce a pair of null vectors $(L,\underline{L})$ with
$$L:=\partial_t+\partial_r\s\s\s\s\mbox{and}\s\s\s\s\underline{L}:=\partial_t-\partial_r,$$
where $x:=(x^0,x^1,x^2,x^3)$, $t:=x^0$ and $r:=\sqrt{(x^1)^2+(x^2)^2+(x^3)^2}$. Sometimes, we also denote $x^{i}$ by $x_{i}$($i=1,2,3$) and $L$ by $\bar{\partial}_0$. Let $\tilde{S}_1$ and $\tilde{S}_2$ be two orthonormal smooth tangent vector fields to the unit sphere $\mathbb{S}^2$, where the orthogonality is in the sense of the standard  metric of $\mathbb{S}^2$. For convenience $A,B,C,D,\cdots$ means any of the vectors $\tilde{S}_1$ and $\tilde{S}_2$ at times. Given a 1-tensor $\pi:=\pi_{\beta}dx^{\beta}$ and a 2-tensor $p:=p_{\alpha\beta}dx^{\alpha}\otimes dx^{\beta}$, we define $\pi_X:=\pi(X)$ and $p_{XY}:=p(X,Y)$, provided $X, Y$ are two vector fields.

Suppose that
$$eu:=dt\otimes dt+\sum\limits_{i=1}^3dx^i\otimes dx^i$$
is the standard Euclidean metric of $\mathbb{R}^4$. Then we have
\begin{eqnarray*}
eu_{\underline{L}L}=eu_{LA}=eu_{\underline{L}A}=0,\s\s\s\s eu_{LL}=eu_{\underline{L}\underline{L}}=2
\end{eqnarray*}
and
\begin{eqnarray*}
eu_{AB}=\delta_{AB}:=\left\{
\begin{array}{rcl}
&0      & A\not=B\\
&1      & A=B.
\end{array} \right.
\end{eqnarray*}
The inverse of $eu$ is
\begin{eqnarray*}
eu^{\underline{L}L}=eu^{LA}=eu^{\underline{L}A}=0,\s\s\s\s eu^{LL}=eu^{\underline{L}\underline{L}}=1/2,\s\s\s\s eu^{AB}=\delta_{AB}.
\end{eqnarray*}

Noting that $\tilde{S}_1$ and $\tilde{S}_2$ are defined only locally on $\mathbb{S}^2$, we replace them with the projections
\begin{eqnarray*}
\bar{\partial}_i:=\partial_i-\omega_i\cdot\partial_r,\s\s\s\s \omega_i:=x_i/r\s\s\mbox{and}\s\s i=1,2,3.
\end{eqnarray*}
It is nor hard to see that $\{\bar{\partial}_1,\bar{\partial}_2,\bar{\partial}_3\}$ gives a set of global and linear dependent vector fields on $\mathbb{S}^2$. Moreover, one can also represent $\{\bar{\partial}_i|i=1,2,3\}$ by $\tilde{S}_1$ and $\tilde{S}_2$, i.e.
\begin{eqnarray}\label{19}
\bar{\partial}_i=\tilde{S}_1^i\cdot \tilde{S}_1+\tilde{S}_2^i\cdot \tilde{S}_2,
\end{eqnarray}
where $\tilde{S}_j:=\tilde{S}_j^i\cdot\partial_i$ and $j=1,2$.

We call $\{L,\underline{L},\tilde{S}_1,\tilde{S}_2\}$ the null frame and introduce the coming notation. Let $\mathcal{T}:=\{L,\tilde{S}_1,\tilde{S}_2\}$, $\mathcal{U}:=\{\underline{L},L,\tilde{S}_1,\tilde{S}_2\}$, $\mathcal{L}:=\{L\}$ and $\mathcal{S}:=\{\tilde{S}_1,\tilde{S}_2\}$. For any $l$ of these families $\mathcal{V}_1,\cdots,\mathcal{V}_l$(namely, $\mathcal{V}_1,\cdots,\mathcal{V}_l\in\{\mathcal{T},\mathcal{U},\mathcal{L},\mathcal{S}\}$) and an arbitrary $k$-tensor $p:=p_{\alpha_1\cdots\alpha_k}dx^{\alpha_1}\otimes\cdots\otimes dx^{\alpha_k}$ with $k\geqslant l$, we define two norms $|p|$ and $|p|_{\mathcal{V}_1\cdots\mathcal{V}_l}$ as
\begin{eqnarray*}
|p|^2:=\sum\limits_{\alpha_1,\cdots,\alpha_k=0}^3(p_{\alpha_1\cdots\alpha_k})^2
\end{eqnarray*}
and
\begin{eqnarray*}
|p|^2_{\mathcal{V}_1\cdots\mathcal{V}_l}:&=&\sum\limits_{V_1,V'_1\in\mathcal{V}_1}\cdots\sum\limits_{V_l,V'_l\in\mathcal{V}_l}\sum\limits_{\alpha_{l+1},\cdots,\alpha_k=0}^3eu^{V_1V_1'}\cdots eu^{V_lV'_l}\\
&&\times p(V_1,\cdots,V_l,\partial_{\alpha_{l+1}},\cdots,\partial_{\alpha_k})\cdot p(V'_1,\cdots,V'_l,\partial_{\alpha_{l+1}},\cdots,\partial_{\alpha_k})
\end{eqnarray*}
It is not difficult to check that $|p|_{\mathcal{V}_1\cdots\mathcal{V}_l}$ is independent of the choice of $\{\tilde{S}_1,\tilde{S}_2\}$ on $\mathbb{S}^2$.

\subsection{The Minkowski metric}
Recall that the Minkowski metric $m$ of $\mathbb{R}^4$ is given by
\begin{eqnarray*}
m:=-dt\otimes dt+\sum\limits_{i=1}^3dx^i\otimes dx^i.
\end{eqnarray*}

From Section 4 of \cite{LR2010} it follows that
\begin{eqnarray*}
m_{LL}=m_{\underline{L}\underline{L}}=m_{LA}=m_{\underline{L}A}=0,\s\s\s\s m_{L\underline{L}}=m_{\underline{L}L}=-2\s\s\s\s m_{AB}=\delta_{AB}
\end{eqnarray*}
The inverse of the metric has the form
\begin{eqnarray*}
m^{LL}=m^{\underline{L}\underline{L}}=m^{LA}=m^{\underline{L}A}=0,\s\s\s\s m^{L\underline{L}}=m^{\underline{L}L}=-1/2,\s\s\s\s m^{AB}=\delta^{AB}.
\end{eqnarray*}

Recall that $\partial$ is the Levi-Civita connection of $(\mathbb{R}^4,m)$. We shall use it to define a new differential operator $\bar{\partial}$ as follows. Provided $p$ is a $k$-tensor and $q:=r-t$, $\bar{\partial}p$ is given by
\begin{eqnarray*}
\bar{\partial}p:=\partial p-\partial_{\partial r}p\otimes dq,
\end{eqnarray*}
where we recall $\partial_r:=\frac{\partial}{\partial r}$. Easily, the readers, reviewing the definition of $\bar{\partial}_{\beta}$ in Subsection \ref{sub2.11}, can check that
\begin{eqnarray*}
\bar{\partial}p=\bar{\partial}_{\beta}\left(p_{\alpha_1\cdots\alpha_k}\right)\cdot dx^{\alpha_1}\otimes\cdots\otimes dx^{\alpha_k}\otimes dx^{\beta}.
\end{eqnarray*}

\begin{rem}
From Lemma 2.6 of \cite{JG} it follows that for any 2-tensor $p$ and $\mathcal{V},\mathcal{W}\in\{\mathcal{T},\mathcal{U},\mathcal{L},\mathcal{S}\}$, the quantity $|\bar{\partial}p|_{\mathcal{V}\mathcal{W}}$ is equivalent to that of (4.5) in \cite{LR2010}. Moreover, $|p|_{\mathcal{V}\mathcal{W}}$ and $|\partial p|_{\mathcal{V}\mathcal{W}}$ are all equivalent to those of (4.3) and (4.4) in \cite{LR2010}.
\end{rem}

\section{Local well-posedness}\label{section3}
Recall that $A$ is the Yang-Mills potential. For simplicity, we decompose $A$ as
$$A:=A^{\mbox{time}}+A^{\mbox{space}},$$
where
$$
A^{\mbox{time}}(x,t):=A_0(x,t)dt\s\s\s\s\mbox{and}\s\s\s\s A^{\mbox{space}}(x,t):=A_i(x,t)dx^i.
$$
From now on, we always assume that $(\mathbb{R}^4,g)$ satisfies the wave coordinates condition.
\subsection{The Yang-Mills equations in wave coordinates and Lorentzian gauges}\label{sub1}
The following computation is obvious
\begin{eqnarray}\label{7}
\nabla^{\alpha}\tilde{F}_{\alpha\beta}&=&g^{\alpha\lambda}(\partial_{\lambda}\partial_{\alpha}A_{\beta}-\partial_{\lambda}\partial_{\beta}A_{\alpha}+[\partial_{\lambda}A_{\alpha},A_{\beta}]+[A_{\alpha},\partial_{\lambda}A_{\beta}])\\
&&-g^{\alpha\lambda}\Gamma^{\theta}_{\lambda\beta}\cdot(\partial_{\alpha}A_{\theta}-\partial_{\theta}A_{\alpha}),\nonumber
\end{eqnarray}
where $\Gamma$ is the Christoffel symbols of $g$ and we have used the wave coordinates condition to deduce (\ref{7}).

By elementary manipulations (\ref{7}) becomes
\begin{eqnarray}\label{8}
\nabla^{\alpha}\tilde{F}_{\alpha\beta}&=&g^{\alpha\lambda}\partial_{\lambda}\partial_{\alpha}A_{\beta}-\partial_{\beta}(g^{\alpha\lambda}\partial_{\lambda}A_{\alpha})+\partial_{\beta}g^{\alpha\lambda}\cdot\partial_{\lambda}A_{\alpha}+g^{\alpha\lambda}[\partial_{\lambda}A_{\alpha},A_{\beta}]+g^{\alpha\lambda}[A_{\alpha},\partial_{\lambda}A_{\beta}]\nonumber\\
&&-g^{\alpha\lambda}\Gamma^{\theta}_{\lambda\beta}\cdot(\partial_{\alpha}A_{\theta}-\partial_{\theta}A_{\alpha}).
\end{eqnarray}
We can transform (\ref{8}) into
\begin{eqnarray*}
\nabla^{\alpha}\tilde{F}_{\alpha\beta}=g^{\alpha\lambda}\partial_{\lambda}\partial_{\alpha}A_{\beta}+\partial_{\beta}g^{\alpha\lambda}\cdot\partial_{\lambda}A_{\alpha}+g^{\alpha\lambda}[A_{\alpha},\partial_{\lambda}A_{\beta}]-g^{\alpha\lambda}\Gamma^{\theta}_{\lambda\beta}\cdot(\partial_{\alpha}A_{\theta}-\partial_{\theta}A_{\alpha}),
\end{eqnarray*}
if we assume that $g^{\alpha\lambda}\partial_{\lambda}A_{\alpha}=0$, which is equivalent to $\mbox{div}_gA\equiv g^{\alpha\lambda}\nabla_{\lambda}A_{\alpha}=0$ called the Lorentz gauge condition(the equivalence follows from the wave gauge condition).

It is easy to see
\begin{eqnarray*}
\partial_{\beta}g^{\alpha\lambda}=-g^{\alpha\theta}\cdot\partial_{\beta}g_{\theta\mu}\cdot g^{\mu\lambda}.
\end{eqnarray*}

Therefore, the Yang-Mills equation in wave coordinates and Lorentzian gauges can be written as
\begin{eqnarray}\label{20}
g^{\alpha\lambda}\partial_{\lambda}\partial_{\alpha}A_{\beta}+f_{\beta}(g,A,\partial g,\partial A,\Phi,\partial\Phi)=0,
\end{eqnarray}
where
\begin{eqnarray*}
&&f_{\beta}(g,A,\partial g,\partial A,\Phi,\partial\Phi):=-g^{\alpha\theta}\cdot\partial_{\beta} g_{\theta\mu}\cdot g^{\mu\lambda}\cdot\partial_{\lambda}A_{\alpha}+g^{\alpha\lambda}[A_{\alpha},\partial_{\lambda}A_{\beta}]+g^{\alpha\lambda}[A_{\lambda},\partial_{\alpha}A_{\beta}]\\
&&-\frac{1}{2}g^{\alpha\lambda}g^{\theta\mu}\cdot(\partial_{\lambda}g_{\mu\beta}+\partial_{\beta}g_{\lambda\mu}-\partial_{\mu}g_{\lambda\beta})\cdot(\partial_{\alpha}A_{\theta}-\partial_{\theta}A_{\alpha})-g^{\alpha\lambda}[A_{\lambda},\partial_{\beta}A_{\alpha}]+g^{\alpha\lambda}[A_{\lambda},[A_{\alpha},A_{\beta}]]\\
&&-\Phi^{\dag}S(\partial_{\beta}\Phi-\sqrt{-1}A^I_{\beta}\sigma_I\Phi/2)+(\partial_{\beta}\Phi-\sqrt{-1}A^I_{\beta}\sigma_I\Phi/2)^{\dag}S\Phi.
\end{eqnarray*}
\subsection{Einstein equation in wave coordinates}\label{sub2}
Referring to Section 7.4 of Chapter 6 in \cite{CB} we get the coming formula
\begin{eqnarray*}
Ric_{\alpha\beta}=-\frac{1}{2}g^{\lambda\mu}\partial_{\lambda}\partial_{\mu}g_{\alpha\beta}+h_{\alpha\beta}(g,\partial g)+\frac{1}{2}(g_{\alpha\lambda}\partial_{\beta}\hat{F}^{\lambda}+g_{\beta\lambda}\partial_{\alpha}\hat{F}^{\lambda})
\end{eqnarray*}
with
\begin{eqnarray*}
h_{\alpha\beta}(g,\partial g):=P_{\alpha\beta}^{\rho\sigma\gamma\delta\lambda\mu}(g,g^{-1})\cdot\partial_{\rho}g_{\gamma\delta}\cdot\partial_{\sigma}g_{\lambda\mu},
\end{eqnarray*}
where $Ric$ is the Ricci tensor of $g$ and the tensor $P$ is a polynomial in $g$ and $g^{-1}$.

Hence, (\ref{0}) can be reduced to
\begin{eqnarray*}
g^{\lambda\mu}\partial_{\lambda}\partial_{\mu}g_{\alpha\beta}+\tilde{f}_{\alpha\beta}(g,A,\Phi,\partial g,\partial A,\partial\Phi)=0,
\end{eqnarray*}
where
\begin{eqnarray*}
&&\tilde{f}_{\alpha\beta}(g,A,\Phi,\partial g,\partial A,\partial\Phi):=2g^{\lambda\mu}\langle\partial_{\lambda}A_{\alpha}-\partial_{\alpha}A_{\lambda}+[A_{\lambda},A_{\alpha}],\partial_{\mu}A_{\beta}-\partial_{\beta}A_{\mu}+[A_{\mu},A_{\beta}]\rangle\\
&&-\frac{1}{2}g_{\alpha\beta}g^{\lambda\rho}g^{\mu\sigma}\langle\partial_{\lambda}A_{\mu}-\partial_{\mu}A_{\lambda}+[A_{\lambda},A_{\mu}],\partial_{\rho}A_{\sigma}-\partial_{\sigma}A_{\rho}+[A_{\rho},A_{\sigma}]\rangle-2h_{\alpha\beta}(g,\partial g)\\
&&-g_{\alpha\lambda}\partial_{\beta}\hat{F}^{\lambda}-g_{\beta\lambda}\partial_{\alpha}\hat{F}^{\lambda}+2(\partial_{\alpha}\Phi-\sqrt{-1}A^{I_1}_{\alpha}\sigma_{I_1}\Phi/2)^{\dag}(\partial_{\beta}\Phi-\sqrt{-1}A^{I_2}_{\beta}\sigma_{I_2}\Phi/2)\\
&&+2(\partial_{\beta}\Phi-\sqrt{-1}A^{I_2}_{\beta}\sigma_{I_2}\Phi/2)^{\dag}(\partial_{\alpha}\Phi-\sqrt{-1}A^{I_1}_{\alpha}\sigma_{I_1}\Phi/2)+2g_{\alpha\beta}\cdot V(|\Phi|^2).
\end{eqnarray*}
The wave coordinates conditions tell us that $\hat{F}^{\lambda}=0$. Hence we have
\begin{eqnarray*}
&&\tilde{f}_{\alpha\beta}(g,A,\Phi,\partial g,\partial A,\partial\Phi):=2g^{\lambda\mu}\langle\partial_{\lambda}A_{\alpha}-\partial_{\alpha}A_{\lambda}+[A_{\lambda},A_{\alpha}],\partial_{\mu}A_{\beta}-\partial_{\beta}A_{\mu}+[A_{\mu},A_{\beta}]\rangle\\
&&-\frac{1}{2}g_{\alpha\beta}g^{\lambda\rho}g^{\mu\sigma}\langle\partial_{\lambda}A_{\mu}-\partial_{\mu}A_{\lambda}+[A_{\lambda},A_{\mu}],\partial_{\rho}A_{\sigma}-\partial_{\sigma}A_{\rho}+[A_{\rho},A_{\sigma}]\rangle-2h_{\alpha\beta}(g,\partial g)\\
&&+2(\partial_{\alpha}\Phi-\sqrt{-1}A^{I_1}_{\alpha}\sigma_{I_1}\Phi/2)^{\dag}(\partial_{\beta}\Phi-\sqrt{-1}A^{I_2}_{\beta}\sigma_{I_2}\Phi/2)+2g_{\alpha\beta}\cdot V(|\Phi|^2)\\
&&+2(\partial_{\beta}\Phi-\sqrt{-1}A^{I_2}_{\beta}\sigma_{I_2}\Phi/2)^{\dag}(\partial_{\alpha}\Phi-\sqrt{-1}A^{I_1}_{\alpha}\sigma_{I_1}\Phi/2).
\end{eqnarray*}

\subsection{Higgs equations in wave coordinates and Lorentzian gauges}\label{sub3.3}
It is easy to check that Higgs equations in wave coordinates and Lorentzian gauges are equivalent to
\begin{eqnarray*}
g^{\lambda\mu}\partial_{\lambda}\partial_{\mu}\Phi+U(\Phi,\partial\Phi,A)=0,
\end{eqnarray*}
where
\begin{eqnarray*}
U(\Phi,\partial\Phi,A):&=&-\frac{\sqrt{-1}}{2}g^{\lambda\mu}A_{\mu}^{I_1}\sigma_{I_1}(\partial_{\lambda}\Phi)-\frac{\sqrt{-1}}{2}g^{\lambda\mu}A_{\lambda}^{I_2}\sigma_{I_2}(\partial_{\mu}\Phi)\\
&&-\frac{1}{4}g^{\lambda\mu}A^{I_2}_{\lambda}A^{I_1}_{\mu}\sigma_{I_2}\sigma_{I_1}(\Phi)-V'(|\Phi|^2)\Phi
\end{eqnarray*}
and the equivalence follows from wave coordinates conditions and Lorentzian gauges.

\subsection{Reducing EYMH equations to quasi-linear systems on a new bundle over $(\mathbb{R}^4,eu)$}\label{sub3.4}
Firstly, we want to construct a new vector bundle $BU$ over $(\mathbb{R}^4,eu)$. It is given by
\begin{eqnarray*}
BU:=(T^*\mathbb{R}^4\otimes T^*\mathbb{R}^4)\times(T^*\mathbb{R}^4\otimes\mathfrak{su}(2))\times\mathbb{C}^2
\end{eqnarray*}
endowed with a metric $\lceil\cdot,\cdot\rceil$, where the symbol ``$\times$'' means the Cartesian product of vector bundles, and $T^*\mathbb{R}^4$ is the cotangent bundle of $\mathbb{R}^4$. More precisely, for any $(g_i,A_i,\Phi_i)\in BU$($i=1,2$), we define
\begin{eqnarray*}
\lceil(g_1,A_1,\Phi_1),(g_2,A_2,\Phi_2)\rceil:=((g_1,g_2))+\langle\langle A_1,A_2\rangle\rangle+\Phi_1^{\dag}\Phi_2,
\end{eqnarray*}
where
\begin{eqnarray*}
((g_1,g_2)):=\sum\limits_{\alpha,\theta}(g_1)_{\alpha\theta}\cdot(g_2)_{\alpha\theta},\s\s\s\s\mbox{and}\s\s\s\s \langle\langle A_1,A_2\rangle\rangle:=\sum\limits_{\alpha}\langle(A_1)_{\alpha},(A_2)_{\alpha}\rangle.
\end{eqnarray*}

Furthermore, we define a connection $\textbf{\mbox{D}}$ on $BU$ by the following identity
\begin{eqnarray*}
\textbf{\mbox{D}}(g,A,\Phi):=(\partial g,\partial A,\partial\Phi)\s\s\s\s\mbox{for any $(g,A,\Phi)\in BU$.}
\end{eqnarray*}
It is not difficult to check that $\textbf{\mbox{D}}$ is compatible to the metric $\lceil\cdot,\cdot\rceil$.

From the discussion in Subsection \ref{sub1}, \ref{sub2} and \ref{sub3.3} we infer that if $u:=(g,A,\Phi)\in BU$ solves the EYMH equations in wave coordinates and Lorentzian gauges, then it is also a solution of the following quasi-linear system
\begin{eqnarray}\label{18}
h^{\lambda\mu}(u)\cdot\textbf{\mbox{D}}_{\lambda}\textbf{\mbox{D}}_{\mu}u+l(u,\textbf{\mbox{D}}u)=0,
\end{eqnarray}
where
\begin{eqnarray*}
h^{\lambda\mu}(u):=g^{\lambda\mu}\s\s\s\s\mbox{and}\s\s\s\s l(u,\textbf{\mbox{D}}u):=\left(\tilde{f}_{\alpha\beta}(u,\textbf{\mbox{D}}u)dx^{\alpha}\otimes dx^{\beta},f_{\theta}(u,\textbf{\mbox{D}}u)dx^{\theta},\hat{U}(u,\textbf{\mbox{D}}u)\right),
\end{eqnarray*}
where $\hat{U}(u,\textbf{\mbox{D}}u):=U(\Phi,\partial\Phi,A)$.

\subsection{Determining the initial value of $\textbf{\mbox{D}}_0u$ on $\mathbb{R}^3$}\label{sub3.5}
In order to determining $\textbf{\mbox{D}}_0u|_{t=0}$, we must give the values of $(\partial_tg_{ij},\partial_t\beta^i,\partial_tN,\partial_tA_{\alpha})|_{t=0}$. From Chapter 6 of \cite{CB} it follows that they can not be chosen arbitrarily; they should satisfy some restrictions.

By (6.1) of Section 6.1 in Chapter 6 of \cite{CB} we know
\begin{eqnarray}\label{15}
\partial_tg_{ij}|_{t=0}=-2\bar{N}K_{ij}+\bar{g}_{jh}\bar{\nabla}_i\bar{\beta}^h+\bar{g}_{ih}\bar{\nabla}_j\bar{\beta}^h.
\end{eqnarray}

From Lorentzian gauge condition $\mbox{div}_gA=0$ and wave coordinates condition we infer that
\begin{eqnarray*}
g^{0\alpha}\partial_tA_{\alpha}=-g^{i\alpha}\partial_iA_{\alpha},
\end{eqnarray*}
implying
\begin{eqnarray*}
-N^{-2}\partial_tA_0+N^{-2}\beta^i\partial_tA_i=-N^{-2}\beta^i\partial_iA_0-(g_*^{ij}-N^{-2}\beta^i\beta^j)\partial_iA_j.
\end{eqnarray*}
Restricting the above identity to $\mathbb{R}^3$ yields
\begin{eqnarray}\label{16}
(\partial_tA_0)|_{t=0}=\bar{\beta}^i(\partial_tA_i)|_{t=0}+\bar{\beta}^i\partial_i\bar{A}_0+(\bar{N}^{2}\bar{g}^{ij}-\bar{\beta}^i\bar{\beta}^j)\partial_i\bar{A}_j,
\end{eqnarray}
where $\bar{A}:=A^{\mbox{space}}|_{t=0}$ and $\bar{A}_0:=A^{\mbox{time}}(\partial_t)|_{t=0}$. Now the problem turns to be how to determine $(\partial_tA_i)|_{t=0}$. Easily, from (\ref{10}) it follows that
\begin{eqnarray}\label{17}
(\partial_tA_i)|_{t=0}=\partial_i\bar{A}_0-[\bar{A}_0,\bar{A}_i]+\bar{N}\cdot\bar{E}^j\cdot\bar{g}_{ji}.
\end{eqnarray}
Substituting (\ref{17}) into (\ref{16}) yields
\begin{eqnarray}\label{24}
(\partial_tA_0)|_{t=0}=2\bar{\beta}^i\partial_i\bar{A}_0-\bar{\beta}^i[\bar{A}_0,\bar{A}_i]+\bar{N}\cdot\bar{g}_{ij}\bar{E}^i\bar{\beta}^j+(\bar{N}^{2}\bar{g}^{ij}-\bar{\beta}^i\bar{\beta}^j)\partial_i\bar{A}_j.
\end{eqnarray}
In other words, if $\bar{A}_0$ and $\bar{A}$ are given, then $(\partial_tA_0)|_{t=0}$ and $(\partial_tA_i)|_{t=0}$ can be specified via (\ref{16}) and (\ref{17}).

The wave coordinates condition tells us
\begin{eqnarray*}
g^{\alpha\beta}\Gamma^0_{\alpha\beta}=0\s\s\s\s\mbox{and}\s\s\s\s g^{\alpha\beta}\Gamma^k_{\alpha\beta}=0,
\end{eqnarray*}
which are equivalent to
\begin{eqnarray}\label{12}
\partial_tN=\frac{1}{2}Ng_*^{ij}\partial_tg_{ij}-N\mbox{div}_{g_t}\beta+\beta(N)
\end{eqnarray}
and
\begin{eqnarray}\label{13}
\partial_t\beta^k&=&(N^2g_*^{ij}-\beta^i\beta^j){}^t\Gamma^k_{ij}-\frac{1}{2}\beta^k\cdot\partial_tg_{ij}\cdot g_*^{ij}+\frac{1}{2}\beta^k\cdot(L_{\beta}g)_{ij}\cdot g_*^{ij}\nonumber\\
&&+N^{-1}\partial_tN\beta^k+\frac{1}{2}g_{*}^{kh}\partial_h\{g(\beta,\beta)-N^2\}-N^{-1}\beta^k\beta(N)\\
&&+\beta^i\cdot\left({}^t\nabla_i\beta^k\right)-\beta^ig_*^{hk}g_{ip}\cdot\left({}^t\nabla_h\beta^p\right).\nonumber
\end{eqnarray}
Substituting (\ref{12}) into (\ref{13}) yields
\begin{eqnarray}\label{14}
\partial_t\beta^k&=&(N^2g_*^{ij}-\beta^i\beta^j){}^t\Gamma^k_{ij}+\frac{1}{2}\beta^k\cdot(L_{\beta}g)_{ij}\cdot g_*^{ij}-\beta^k\mbox{div}_{g_t}\beta\nonumber\\
&&+\frac{1}{2}g_{*}^{kh}\partial_h\{g(\beta,\beta)-N^2\}+\beta^i\cdot\left({}^t\nabla_i\beta^k\right)-\beta^ig_*^{hk}g_{ip}\cdot\left({}^t\nabla_h\beta^p\right).
\end{eqnarray}
Restricting (\ref{12}) and (\ref{13}) to $\mathbb{R}^3$ and then substituting (\ref{15}) into them lead to
\begin{eqnarray}\label{25}
\partial_tN|_{t=0}=-\bar{N}^2\cdot\mbox{tr}_{\bar{g}}K+\bar{\beta}(\bar{N})
\end{eqnarray}
and
\begin{eqnarray}\label{26}
\partial_t\beta^k|_{t=0}&=&(\bar{N}^2\bar{g}^{ij}-\bar{\beta}^i\bar{\beta}^j)\bar{\Gamma}^k_{ij}+\frac{1}{2}\bar{\beta}^k\cdot\mbox{tr}_{\bar{g}}(L_{\bar{\beta}}\bar{g})-\bar{\beta}^k\mbox{div}_{\bar{g}}\bar{\beta}\\
&&+\frac{1}{2}\bar{g}^{kh}\partial_h\{\bar{g}(\bar{\beta},\bar{\beta})-\bar{N}^2\}+\bar{\beta}^i\cdot\bar{\nabla}_i\bar{\beta}^k-\bar{\beta}^i\bar{g}^{hk}\bar{g}_{ip}\cdot\bar{\nabla}_h\bar{\beta}^p.\nonumber
\end{eqnarray}

\subsection{Local existence and uniqueness in the wave coordinates and Lorentzian gauge}
Thanks to Subsection \ref{sub3.4} and \ref{sub3.5}, we have formulated the intrinsic Cauchy problem for EYMH equations in the form of standard PDE analyses. Hence, one can now use the results in Appendix 3 of \cite{CB} to obtain a local in time, global in space, existence and uniqueness theorem in the wave coordinates and Lorentzian gauges. The methods we rely on are almost the same as those of Section 7 and 8 in Chapter 6 of \cite{CB}. Before getting the local existence and uniqueness theorem, we need two lemmas.

\begin{lem}\label{lem3.1}
If $(g,A,\Phi)$ satisfies the EYMH equations in the wave coordinates and Lorentzian gauge, then the wave functions $\hat{F}^{\lambda}$ and the function $\mbox{div}_gA$ satisfy a system of second order and linear homogeneous differential equations with principal terms the wave equation in the metric $g$.
\end{lem}

\textbf{Proof.} It is easy if the readers apply Bianchi identities. The process of proof is almost the same as that of Lemma 10.1 in Chapter 6 of \cite{CB}. Hence we omit it.
\endproof

\begin{lem}\label{lem3.2}
Given a solution of the EYMH equations in wave coordinates and Lorentzian gauge, whose initial datum satisfy $\hat{F}^{\lambda}|_{t=0}=0$ and $\mbox{div}_gA|_{t=0}=0$, the conditions $\partial_t\hat{F}^{\lambda}|_{t=0}=0$ and $\partial_t(\mbox{div}_gA)|_{t=0}=0$ are satisfied if and only if the initial datum satisfy the EYMH constraints.
\end{lem}

\textbf{Proof.} The result follows from straightforward computation. \endproof

\begin{thm}
Let $e$ be the standard Euclidean metric of $\mathbb{R}^3$ and $\tilde{D}$ is the Levi-Civita connection of $(\mathbb{R}^3,e)$.

Hypotheses on the initial datum sets $(\mathbb{R}^3,\bar{g},K,\bar{A}_0,\bar{A}^{\mbox{space}},\bar{E},\bar{\beta},\bar{N},\bar{\Phi},\bar{\Psi})$ and\\$(\partial_tg_{ij}|_{t=0}, \partial_tA_0|_{t=0}, \partial_tA_i|_{t=0}, \partial_t\beta^k|_{t=0}, \partial_tN|_{t=0})$:

1. $\bar{g}$ is a Riemannian metric on $\mathbb{R}^3$ uniformly equivalent to $e$ and such that
\begin{eqnarray*}
\tilde{D}\bar{g}\in H_{s-1}\s\s\s\s\mbox{and}\s\s\s\s\bar{g}\in\bar{C}^0\s\s\s\s\mbox{with}\s\s\s\s s\in\mathbb{Z}\cap\left[3,\infty\right),
\end{eqnarray*}
where $\mathbb{Z}$ is the set of all the integers. Furthermore, $(\partial_tg_{ij})|_{t=0}$ is given by (\ref{15}).

2. $K$ is a symmetric 2-tensor on $\mathbb{R}^3$ such that $K\in H_{s-1}.$

3. $\bar{A}_0$ belongs to $\bar{C}^0$ and $\partial\bar{A}_0\in H_{s-1}$. And $(\partial_tA_0)|_{t=0}$ is given by (\ref{24}).

4. $\bar{A}^{\mbox{space}}\in\bar{C}^0$ and $\tilde{D}(\bar{A}^{\mbox{space}})\in H_{s-1}$. Moreover, $(\partial_tA_i)|_{t=0}$ is given by (\ref{17}).

5. $\bar{\beta}\in\bar{C}^0$ and $\tilde{D}\bar{\beta}\in H_{s-1}$. Moreover, there exists a positive constant $b$ such that $e(\bar{\beta},\bar{\beta})\leqslant b$. And $\partial_t\beta^k|_{t=0}$ is given by (\ref{26}).

6. $\bar{N}\in\bar{C}^0$ and $\partial\bar{N}\in H_{s-1}$. Moreover, there exist two positive constants $C_1$ and $C_2$ such that $C_1\leqslant\bar{N}\leqslant C_2$. Besides, $\partial_tN|_{t=0}$ is given by (\ref{25}).

7. $\bar{E}\in H_{s-1}$.

8. $\bar{\Phi}\in\bar{C}^0$ and $\partial\bar{\Phi}\in H_{s-1}$.

9. $\bar{\Psi}\in H_{s-1}$.

10. $(\mathbb{R}^3,\bar{g},K,\bar{A}_0,\bar{A}^{\mbox{space}},\bar{E},\bar{\beta},\bar{N},\bar{\Phi},\bar{\Psi})$ satisfies the EYMH constraints.

Conclusions:

The initial datum sets admit a development $(V_T,g,A,\Phi)$ for some $T>0$, such that $A\in\tilde{E}_s(T)$, the spacetime metric $g$ is a regular sliced Lorentzian metric in $\tilde{E}_s(T)$, $\Phi\in\tilde{E}_s(T)$ and $(g,A,\Phi)$ satisfies on $V_T$ the EYMH equations. Furthermore, $(g,A,\Phi)$ meets the wave coordinates and Lorentzian gauge conditions.

Two such developments in the wave coordinates and Lorentzian gauge $(V_T,g_1,A_1,\Phi_1)$ and $(V_T,g_2,A_2,\Phi_2)$, which are in $\tilde{E}_s(T)$, and which take the same initial values\\ $(\bar{g},K,\bar{A}_0,\bar{A}^{\mbox{space}},\bar{E},\bar{\beta},\bar{N},\bar{\Phi},\bar{\Psi})$ on $\mathbb{R}^3$, coincide on $V_T$.
\end{thm}

\textbf{Sketch of the proof.}
Note that (\ref{18}) are quasi-diagonal, hyperquasi-linear(i.e. $h^{\lambda\mu}$ depends on $u$ but not on $\textbf{\mbox{D}}u$), second order systems of the type treated in Appendix 3 of \cite{CB}. They satisfies the hypotheses enunciated in that appendix. So the existence and uniqueness theorem for (\ref{18}) then follows.

By Lemma \ref{lem3.2} we know that, since the initial datum satisfy the EYMH constraints and
\begin{eqnarray}\label{27}
\hat{F}^{\lambda}|_{t=0}=0,\s\s\s\s (\mbox{div}_gA)|_{t=0}=0,
\end{eqnarray}
the following identities hold true
\begin{eqnarray*}
\partial_t\hat{F}^{\lambda}|_{t=0}=0\s\s\mbox{and}\s\s\partial_t(\mbox{div}_gA)|_{t=0}=0,
\end{eqnarray*}
where it is obvious that the conditions (\ref{17}), (\ref{24}), (\ref{15}), (\ref{25}) and (\ref{26}) lead to (\ref{27}). Furthermore, Lemma \ref{lem3.1} tells us that, if $(g,A,\Phi)$ satisfies (\ref{18}), then $\hat{F}^{\lambda}$ and $\mbox{div}_gA$ satisfy a system of second order linear homogeneous differential equations with principal terms the wave equation in the metric $g$. Combining the above two lemmas we arrive at that
$$\hat{F}^{\lambda}=0\s\s\s\s\mbox{and}\s\s\s\s\mbox{div}_gA=0,$$
provided the initial datum satisfy the EYMH constraints and (\ref{27}). Hence, a solution for (\ref{18}), with initial datum satisfying the EYMH constraints and (\ref{27}), is a solution for the full EYMH system.
\endproof

\section{The equations of $(h:=g-m,A,\Phi)$}\label{section4}
Given the initial datum set $(\mathbb{R}^3;\bar{g},K,\bar{A}^{\mbox{space}},\bar{A}_0,\bar{E},\bar{\beta}\equiv0,\bar{N},\bar{\Phi},\bar{\Psi})$ satisfying the EYMH constraints:
\begin{eqnarray*}
\bar{R}-|K|^2_{\bar{g}}+(\mbox{tr}_{\bar{g}}K)^2&=&\bar{g}^{ik}\bar{g}^{jl}\langle\bar{F}_{ij},\bar{F}_{kl}\rangle/2+3\bar{g}_{ij}\langle\bar{E}^i,\bar{E}^j\rangle+2|\bar{\Psi}|^2/\bar{N}^2\\
&&+2\{\bar{g}^{ij}(\hat{\rlap{/}\nabla}_i\bar{\Phi})^{\dag}(\hat{\rlap{/}\nabla}_j\bar{\Phi})+V(|\bar{\Phi}|^2)\}
\end{eqnarray*}
with $\bar{F}:=d\bar{A}^{\mbox{space}}+[\bar{A}^{\mbox{space}},\bar{A}^{\mbox{space}}]$,
\begin{eqnarray*}
(\mbox{div}_{\bar{g}}K)_i-\partial_i(\mbox{tr}_{\bar{g}}K)=\langle\bar{E}^l,\bar{F}_{li}\rangle-\bar{N}^{-1}(\bar{\Psi})^{\dag}\hat{\rlap{/}\nabla}_i\bar{\Phi}-\bar{N}^{-1}(\hat{\rlap{/}\nabla}_i\bar{\Phi})^{\dag}\bar{\Psi}
\end{eqnarray*}
and
\begin{eqnarray*}
\bar{\Psi}^{\dag}S\bar{\Phi}-\bar{\Phi}^{\dag}S\bar{\Psi}=\bar{N}\cdot[\bar{A}^{\mbox{space}}_i,\bar{E}^i]+\bar{E}(\bar{N})+\bar{N}\mbox{div}_{\bar{g}}\bar{E},
\end{eqnarray*}
we are going to get a solution for the EYMH equations. Suppose
\begin{eqnarray*}
\bar{g}\in H_{N+1},\s\s K\in H_{N+1},\s\s\bar{E}\in H_{N+1},\s\s\bar{N}\in H_{N+1},
\end{eqnarray*}
\begin{eqnarray*}
\bar{A}^{\mbox{space}}\in H_{N+1},\s\s\bar{A}_0\in H_{N+1},\s\s\bar{\Phi}\in H_{N+1},\s\s\bar{\Psi}\in H_N,
\end{eqnarray*}
where the ``$N$" in ``$H_{N+1}$" is the same as that in ``$\mathcal{E}_N$" and is an integer not smaller than 4. Furthermore, we have to assume that $\bar{g}$ is uniformly equivalent to $e$ and $\bar{N}$ is bounded above and below by some positive constants.

In order to satisfy the wave coordinates and Lorentzian gauge conditions, we define the initial datum $\Phi|_{t=0}$, $\partial_t\Phi|_{t=0}$, $g_{\mu\nu}|_{t=0}$, $\partial_tg_{\mu\nu}|_{t=0}$, $A_{\alpha}|_{t=0}$, and $\partial_tA_{\alpha}|_{t=0}$ as follows:
\begin{eqnarray}\label{28}
g_{ij}|_{t=0}:=\bar{g}_{ij},\s\s g_{00}|_{t=0}:=-\bar{N}^2,\s\s g_{0i}|_{t=0}:=0,
\end{eqnarray}
\begin{eqnarray}\label{29}
A^{\mbox{space}}|_{t=0}:=\bar{A}^{\mbox{space}},\s\s A^{\mbox{time}}|_{t=0}:=\bar{A}_0dt|_{t=0},\s\s\Phi|_{t=0}:=\bar{\Phi},
\end{eqnarray}
\begin{eqnarray}\label{30}
&&\partial_tg_{ij}|_{t=0}:=-2\bar{N}K_{ij},\s\s\partial_tg_{00}|_{t=0}:=2\bar{N}^3\cdot\mbox{tr}_{\bar{g}}K,
\end{eqnarray}
\begin{eqnarray}\label{31}
\partial_tg_{0l}|_{t=0}:=\bar{N}^2\bar{g}^{ij}\partial_j\bar{g}_{il}-\frac{1}{2}\bar{N}^2\bar{g}^{ij}\partial_l\bar{g}_{ij}-\bar{N}\partial_l\bar{N},
\end{eqnarray}
\begin{eqnarray}\label{32}
\partial_tA_0|_{t=0}:=\bar{N}^2\bar{g}^{ij}\partial_i\bar{A}^{\mbox{space}}_j,\s\partial_tA_i|_{t=0}:=\partial_i\bar{A}_0-[\bar{A}_0,\bar{A}^{\mbox{space}}_i]+\bar{N}\cdot\bar{E}^j\cdot\bar{g}_{ji},
\end{eqnarray}
\begin{eqnarray}\label{107}
\partial_t\Phi|_{t=0}:=\bar{\Psi}+\sqrt{-1}(\bar{A}_0)^I\sigma_I\bar{\Phi}/2.
\end{eqnarray}
From (\ref{11}) it follows that giving $\partial_tg_{0l}|_{t=0}$ and $\partial_tg_{00}|_{t=0}$ is equivalent to giving $\partial_t\beta^k|_{t=0}$ and $\partial_tN|_{t=0}$.

Now we obtain a solution $(g,A,\Phi)\in\tilde{E}_{N+1}(T)$ to the EYMH equations for some $T>0$, which also satisfies the wave coordinates and Lorentzian gauge conditions.

On the other hand, from (3.17) of \cite{LR2005} it follows that
\begin{eqnarray*}
Ric_{\mu\nu}=-\frac{1}{2}\overset{\sim}{\Box}_gg_{\mu\nu}+\frac{1}{2}\tilde{P}(\partial_\mu g,\partial_{\nu} g)+\frac{1}{2}\tilde{Q}_{\mu\nu}(\partial g,\partial g),
\end{eqnarray*}
where
\begin{eqnarray*}
\tilde{P}(\partial_\mu g,\partial_{\nu} g):=g^{\alpha\alpha'}g^{\beta\beta'}\cdot\left(\frac{1}{4}\partial_{\mu}g_{\beta\beta'}\partial_{\nu}g_{\alpha\alpha'}-\frac{1}{2}\partial_{\nu}g_{\alpha\beta}\partial_{\mu}g_{\alpha'\beta'}\right)
\end{eqnarray*}
and
\begin{eqnarray*}
&&\tilde{Q}_{\mu\nu}(\partial g,\partial g):=g^{\alpha\alpha'}g^{\beta\beta'}\partial_{\alpha}g_{\beta\mu}\partial_{\alpha'}g_{\beta'\nu}-g^{\alpha\alpha'}g^{\beta\beta'}\left(\partial_{\alpha}g_{\beta\mu}\partial_{\beta'}g_{\alpha'\nu}-\partial_{\beta'}g_{\beta\mu}\partial_{\alpha}g_{\alpha'\nu}\right)\\
&&+g^{\alpha\alpha'}g^{\beta\beta'}(\partial_{\mu}g_{\alpha'\beta'}\partial_{\alpha}g_{\beta\nu}-\partial_{\alpha}g_{\alpha'\beta'}\partial_{\mu}g_{\beta\nu})+g^{\alpha\alpha'}g^{\beta\beta'}(\partial_{\nu}g_{\alpha'\beta'}\partial_{\alpha}g_{\beta\mu}-\partial_{\alpha}g_{\alpha'\beta'}\partial_{\nu}g_{\beta\mu})\\
&&+\frac{1}{2}g^{\alpha\alpha'}g^{\beta\beta'}(\partial_{\beta'}g_{\alpha\alpha'}\partial_{\mu}g_{\beta\nu}-\partial_{\mu}g_{\alpha\alpha'}\partial_{\beta'}g_{\beta\nu})+\frac{1}{2}g^{\alpha\alpha'}g^{\beta\beta'}(\partial_{\beta'}g_{\alpha\alpha'}\partial_{\nu}g_{\beta\mu}-\partial_{\nu}g_{\alpha\alpha'}\partial_{\beta'}g_{\beta\mu}).
\end{eqnarray*}
Hence, (\ref{0}) is equivalent to
\begin{eqnarray}\label{22}
\overset{\sim}{\Box}_gg_{\alpha\beta}&=&-2g^{\mu\lambda}\langle\partial_{\alpha}A_{\mu}-\partial_{\mu}A_{\alpha}+[A_{\alpha},A_{\mu}],\partial_{\beta}A_{\lambda}-\partial_{\lambda}A_{\beta}+[A_{\beta},A_{\lambda}]\rangle\nonumber\\
&&+\frac{1}{2}g_{\alpha\beta}g^{\rho\sigma}g^{\mu\lambda}\langle\partial_{\rho}A_{\mu}-\partial_{\mu}A_{\rho}+[A_{\rho},A_{\mu}],\partial_{\sigma}A_{\lambda}-\partial_{\lambda}A_{\sigma}+[A_{\sigma},A_{\lambda}]\rangle\\
&&+\tilde{P}(\partial_{\alpha} g,\partial_{\beta} g)+\tilde{Q}_{\alpha\beta}(\partial g,\partial g)-(\hat{\nabla}_{\alpha}\Phi)^{\dag}\hat{\nabla}_{\beta}\Phi-(\hat{\nabla}_{\beta}\Phi)^{\dag}\hat{\nabla}_{\alpha}\Phi-g_{\alpha\beta}V(|\Phi|^2).\nonumber
\end{eqnarray}

Define two 2-tensors
\begin{eqnarray*}
h_{\mu\nu}:=g_{\mu\nu}-m_{\mu\nu}\s\s\s\s\mbox{and}\s\s\s\s H^{\mu\nu}:=g^{\mu\nu}-m^{\mu\nu},
\end{eqnarray*}
where $m^{\mu\nu}$ and $g^{\mu\nu}$ are the inverses of $m_{\mu\nu}$ and $g_{\mu\nu}$ respectively(Recall that $m_{\mu\nu}$ is the Minkowski metric of $\mathbb{R}^4$). We want to obtain the equation of $h_{\mu\nu}$. From Lemma 4.1 of \cite{JG} it follows that
\begin{eqnarray}\label{YL}
H^{\mu\nu}=-h^{\mu\nu}+O^{\mu\nu}(h^2),
\end{eqnarray}
where $O^{\mu\nu}(h^2)$ is a two-tensor vanishing to the second order at $h=0$. Besides, by Lemma 3.2 of \cite{LR2005}, we know that if $h$ is small, (\ref{22}) is equivalent to
\begin{eqnarray}\label{23}
\overset{\sim}{\Box}_gh_{\mu\nu}&=&P(\partial_{\mu}h,\partial_{\nu}h)+Q_{\mu\nu}(\partial h,\partial h)+G_{\mu\nu}(h)(\partial h,\partial h)-(h_{\mu\nu}+m_{\mu\nu})V(|\Phi|^2)\nonumber\\
&&+2(h^{\alpha\beta}-m^{\alpha\beta})\langle\partial_{\alpha}A_{\mu}-\partial_{\mu}A_{\alpha}+[A_{\alpha},A_{\mu}],\partial_{\beta}A_{\nu}-\partial_{\nu}A_{\beta}+[A_{\beta},A_{\nu}]\rangle\\
&&-2O^{\alpha\beta}(h^2)\langle\partial_{\alpha}A_{\mu}-\partial_{\mu}A_{\alpha}+[A_{\alpha},A_{\mu}],\partial_{\beta}A_{\nu}-\partial_{\nu}A_{\beta}+[A_{\beta},A_{\nu}]\rangle\nonumber\\
&&+\frac{1}{2}\{m^{\alpha\rho}m^{\beta\sigma}h_{\mu\nu}+m^{\alpha\rho}m^{\beta\sigma}m_{\mu\nu}-h^{\alpha\rho}m^{\beta\sigma}m_{\mu\nu}-m^{\alpha\rho}h^{\beta\sigma}m_{\mu\nu}\nonumber\\
&&+O^{\alpha\rho\beta\sigma}_{\mu\nu}(h^2)\}\langle\partial_{\alpha}A_{\beta}-\partial_{\beta}A_{\alpha}+[A_{\alpha},A_{\beta}],\partial_{\rho}A_{\sigma}-\partial_{\sigma}A_{\rho}+[A_{\rho},A_{\sigma}]\rangle\nonumber\\
&&-(\partial_{\mu}\Phi-\sqrt{-1}A^{I_1}_{\mu}\sigma_{I_1}\Phi/2)^{\dag}(\partial_{\nu}\Phi-\sqrt{-1}A^{I_2}_{\nu}\sigma_{I_2}\Phi/2)\nonumber\\
&&-(\partial_{\nu}\Phi-\sqrt{-1}A^{I_2}_{\nu}\sigma_{I_2}\Phi/2)^{\dag}(\partial_{\mu}\Phi-\sqrt{-1}A^{I_1}_{\mu}\sigma_{I_1}\Phi/2)\nonumber\\
&:=&F_{\mu\nu}\nonumber
\end{eqnarray}
where $O^{\alpha\rho\beta\sigma}_{\mu\nu}(h^2)$ vanishes to the second order at $h=0$,
\begin{eqnarray*}
P(\partial_{\mu}h,\partial_{\nu}h):=m^{\alpha\alpha'}m^{\beta\beta'}\cdot\left(\frac{1}{4}\partial_{\mu}h_{\beta\beta'}\partial_{\nu}h_{\alpha\alpha'}-\frac{1}{2}\partial_{\nu}h_{\alpha\beta}\partial_{\mu}h_{\alpha'\beta'}\right)
\end{eqnarray*}
and
\begin{eqnarray*}
&&Q_{\mu\nu}(\partial h,\partial h):=m^{\alpha\alpha'}m^{\beta\beta'}\partial_{\alpha}h_{\beta\mu}\partial_{\alpha'}h_{\beta'\nu}-m^{\alpha\alpha'}m^{\beta\beta'}\left(\partial_{\alpha}h_{\beta\mu}\partial_{\beta'}h_{\alpha'\nu}-\partial_{\beta'}h_{\beta\mu}\partial_{\alpha}h_{\alpha'\nu}\right)\\
&&+m^{\alpha\alpha'}m^{\beta\beta'}(\partial_{\mu}h_{\alpha'\beta'}\partial_{\alpha}h_{\beta\nu}-\partial_{\alpha}h_{\alpha'\beta'}\partial_{\mu}h_{\beta\nu})+m^{\alpha\alpha'}m^{\beta\beta'}(\partial_{\nu}h_{\alpha'\beta'}\partial_{\alpha}h_{\beta\mu}-\partial_{\alpha}h_{\alpha'\beta'}\partial_{\nu}h_{\beta\mu})\\
&&+\frac{1}{2}m^{\alpha\alpha'}m^{\beta\beta'}(\partial_{\beta'}h_{\alpha\alpha'}\partial_{\mu}h_{\beta\nu}-\partial_{\mu}h_{\alpha\alpha'}\partial_{\beta'}h_{\beta\nu})+\frac{1}{2}m^{\alpha\alpha'}m^{\beta\beta'}(\partial_{\beta'}h_{\alpha\alpha'}\partial_{\nu}h_{\beta\mu}-\partial_{\nu}h_{\alpha\alpha'}\partial_{\beta'}h_{\beta\mu}),
\end{eqnarray*}
is a null form and $G_{\mu\nu}(h)(\partial h,\partial h)$ is a quadratic form in $\partial h$ with coefficients smoothly dependent on $h$ and vanishing when $h$ vanishes, i.e. $G_{\mu\nu}(0)(\partial h,\partial h)=0$.

Using (\ref{YL}) again we get
\begin{eqnarray}\label{36}
&&\overset{\sim}{\Box}_gA_{\beta}=\{-m^{\alpha\theta}h^{\mu\lambda}-h^{\alpha\theta}m^{\mu\lambda}+m^{\alpha\theta}m^{\mu\lambda}+O^{\alpha\theta\mu\lambda}(h^2)\}\partial_{\beta}h_{\theta\mu}\partial_{\lambda}A_{\alpha}\\
&&+\frac{1}{2}\{-m^{\alpha\lambda}h^{\theta\mu}-h^{\alpha\lambda}m^{\theta\mu}+m^{\alpha\lambda}m^{\theta\mu}+O^{\alpha\lambda\theta\mu}(h^2)\}\nonumber\\
&&\times(\partial_{\lambda}h_{\mu\beta}+\partial_{\beta}h_{\mu\lambda}-\partial_{\mu}h_{\lambda\beta})(\partial_{\alpha}A_{\theta}-\partial_{\theta}A_{\alpha})-(m^{\alpha\lambda}-h^{\alpha\lambda}+O^{\alpha\lambda}(h^2))[A_{\alpha},\partial_{\lambda}A_{\beta}]\nonumber\\
&&+(m^{\alpha\lambda}-h^{\alpha\lambda}+O^{\alpha\lambda}(h^2))\{[A_{\lambda},\partial_{\beta}A_{\alpha}]-[A_{\lambda},\partial_{\alpha}A_{\beta}]-[A_{\lambda},[A_{\alpha},A_{\beta}]]\}\nonumber\\
&&+\Phi^{\dag}S(\partial_{\beta}\Phi-\sqrt{-1}A^I_{\beta}\sigma_I\Phi/2)-(\partial_{\beta}\Phi-\sqrt{-1}A^I_{\beta}\sigma_I\Phi/2)^{\dag}S\Phi\nonumber\\
&&:=J_{\beta},\nonumber
\end{eqnarray}
provided $h$ is sufficiently small. Easily, the coming identity follows again from (\ref{YL})
\begin{eqnarray}\label{99}
\overset{\sim}{\Box}_g\Phi&=&\frac{\sqrt{-1}}{2}(m^{\lambda\mu}-h^{\lambda\mu}+O^{\lambda\mu}(h^2))A_{\mu}^{I_1}\sigma_{I_1}(\partial_{\lambda}\Phi)+\frac{\sqrt{-1}}{2}(m^{\lambda\mu}-h^{\lambda\mu}+O^{\lambda\mu}(h^2))A_{\lambda}^{I_2}\sigma_{I_2}(\partial_{\mu}\Phi)\nonumber\\
&&\s\s\s\s+\frac{1}{4}(m^{\lambda\mu}-h^{\lambda\mu}+O^{\lambda\mu}(h^2))A^{I_2}_{\lambda}A^{I_1}_{\mu}\sigma_{I_2}\sigma_{I_1}(\Phi)+V'(|\Phi|^2)\Phi:=W.
\end{eqnarray}

\section{Beginning of the proof of Theorem \ref{thm6.1}}\label{section5}
As described in the introduction, $T$ is the maximal existence time of the solution $(g,A,\Phi)$ and assumed to be finite. We have defined the time $T^*$ to be
\begin{eqnarray}\label{38}
T^*:=\sup\{T_0:\exists C=C(T),\,s.t.\s\forall t\in[0,T_0),\s\mathcal{E}_N(t)\leqslant2C\varepsilon\},
\end{eqnarray}
where $\mathcal{E}_N$ is given by (\ref{95}). Our goal is to show that if $\varepsilon>0$ is small enough, then the inequality in (\ref{38}) implies the same inequality with $2C$ replaced by $C$ for all $t<T^*$.

The first step is to derive the preliminary decay estimates for $h$, $A$ and $\Phi$ under the assumption (\ref{38}). However, our method is the same as that of Theorem 5.2 of \cite{JG}. Hence we only list the result and omit the proof.
\begin{thm}\label{thm6.2}
Let $h$, $A$ and $\Phi$ verify the inequality in (\ref{38}). Then we have
\begin{equation}\label{39}
|\partial Z^Ih(x,t)|+|\partial Z^IA(x,t)|+|\partial Z^I\Phi(x,t)|\lesssim\left\{
\begin{array}{rcl}
\varepsilon(1+|q|)^{-2-\gamma}&        \text{$q>0$,}&\\
\varepsilon(1+|q|)^{-3/2}&         \text{$q<0$,}&
\end{array} \right.
|I|\leqslant N-3.
\end{equation}
Furthermore
\begin{equation}\label{40}
|Z^Ih(x,t)|+|Z^IA(x,t)|+|Z^I\Phi(x,t)|\lesssim
\left\{
\begin{array}{rcl}
\varepsilon(1+|q|)^{-1-\gamma}&        \text{$q>0$,}&\\
\varepsilon(1+|q|)^{-1/2}&         \text{$q<0$,}&
\end{array} \right.
|I|\leqslant N-3.
\end{equation}
And
\begin{equation}\label{41}
|\bar{\partial} Z^Ih(x,t)|+|\bar{\partial} Z^IA(x,t)|+|\bar{\partial} Z^I\Phi(x,t)|\lesssim
\left\{
\begin{array}{rcl}
\varepsilon(1+|q|)^{-2-\gamma}&        \text{$q>0$,}&\\
\varepsilon(1+|q|)^{-3/2}&         \text{$q<0$,}&
\end{array} \right.
|I|\leqslant N-4.
\end{equation}
\end{thm}

\subsection{Estimates for the inhomogeneous terms $F_{\mu\nu}$, $J_{\beta}$ and $W$}
(\ref{40}) tells us that $|Z^Ih|+|Z^IA|+|Z^I\Phi|\leqslant1/2$, provided $\varepsilon$ is small enough and $|I|\leqslant N-3$. The upper bound ``1/2" plays a key role in the sequel.
\begin{pro}\label{pro6.3}
Assume that $h=g-m$, $A$ and $\Phi$ satisfy the inequality in (\ref{38}). Let $F_{\mu\nu}$, $J_{\beta}$ and $W$ be as in (\ref{23}), (\ref{36}) and (\ref{99}) respectively. Then we have
\begin{eqnarray}\label{47}
|Z^IF|&\lesssim&\sum\limits_{|J|+|K|\leqslant|I|}(|\partial Z^Jh|_{\mathcal{T}\mathcal{U}}\cdot|\partial Z^Kh|_{\mathcal{T}\mathcal{U}}+|\bar{\partial}Z^Jh|\cdot|\partial Z^Kh|)\nonumber\\
&&+\sum\limits_{|J|+|K|\leqslant|I|-1}|\partial Z^Jh|_{\mathcal{L}\mathcal{T}}\cdot|\partial Z^Kh|+\sum\limits_{|J|+|K|\leqslant|I|-2}|\partial Z^Jh|\cdot|\partial Z^Kh|\\
&&+\sum\limits_{|J_1|+|J_2|+|J_3|\leqslant|I|}|Z^{J_3}h|\cdot|\partial Z^{J_2}h|\cdot|\partial Z^{J_1}h|+\sum\limits_{|J_1|+|J_2|\leqslant|I|}|\partial Z^{J_1}A|\cdot|\partial Z^{J_2}A|\nonumber\\
&&+\sum\limits_{|J_1|+|J_2|+|J_3|\leqslant|I|}|\partial Z^{J_1}A|\cdot|Z^{J_2}A|\cdot|Z^{J_3}A|+\sum\limits_{|I_1|+\cdots+|I_{2s}|=|I|}|V^{(s)}(|\Phi|^2)|\cdot\prod\limits_{i=1}^{2s}|Z^{I_i}\Phi|\nonumber\\
&&+\sum\limits_{|J_1|+|J_2|+|J_3|+|J_4|\leqslant|I|}|Z^{J_1}A|\cdot|Z^{J_2}A|\cdot|Z^{J_3}A|\cdot|Z^{J_4}A|\nonumber\\
&&+\sum\limits_{|J_1|+|I_1|+\cdots+|I_{2s}|=|I|}|Z^{J_1}h|\cdot|V^{(s)}(|\Phi|^2)|\cdot\prod\limits_{i=1}^{2s}|Z^{I_i}\Phi|\nonumber\\
&&+\sum\limits_{|J_1|+|J_2|\leqslant|I|}|\partial Z^{J_1}\Phi|\cdot|\partial Z^{J_2}\Phi|+\sum\limits_{|J_1|+|J_2|+|J_3|\leqslant|I|}|Z^{J_1}A|\cdot|Z^{J_2}\Phi|\cdot|\partial Z^{J_3}\Phi|\nonumber\\
&&+\sum\limits_{|J_1|+|J_2|+|J_3|+|J_4|=|I|}|Z^{J_1}A|\cdot|Z^{J_2}A|\cdot|Z^{J_3}\Phi|\cdot|Z^{J_4}\Phi|,\nonumber
\end{eqnarray}
\begin{eqnarray}\label{48}
|Z^IJ|&\lesssim&\sum\limits_{|I_1|+|I_2|\leqslant|I|}|\partial Z^{I_1}h|\cdot|\partial Z^{I_2}A|+\sum\limits_{|I_1|+|I_2|\leqslant|I|}| Z^{I_1}A|\cdot|\partial Z^{I_2}A|\nonumber\\
&&+\sum\limits_{|I_1|+|I_2|+|I_3|\leqslant|I|}|Z^{I_1}A|\cdot|Z^{I_2}A|\cdot|Z^{I_3}A|+\sum\limits_{|I_1|+|I_2|\leqslant|I|}| Z^{I_1}\Phi|\cdot|\partial Z^{I_2}\Phi|\\
&&+\sum\limits_{|I_1|+|I_2|+|I_3|=|I|}|Z^{I_1}\Phi|\cdot|Z^{I_2}\Phi|\cdot|Z^{I_3}A|\nonumber
\end{eqnarray}
and
\begin{eqnarray}\label{100}
|Z^IW|&\lesssim&\sum\limits_{|I_1|+|I_2|\leqslant|I|}|Z^{I_1}A|\cdot|\partial Z^{I_2}\Phi|+\sum\limits_{|I_1|+|I_2|+|I_3|\leqslant|I|}|Z^{I_1}A|\cdot|Z^{I_2}A|\cdot|Z^{I_3}\Phi|\nonumber\\
&&+\sum\limits_{|I_1|+\cdots+|I_{2s+1}|=|I|}|V^{(s+1)}(|\Phi|^2)|\cdot\prod\limits_{i=1}^{2s+1}|Z^{I_i}\Phi|,
\end{eqnarray}
provided $|Z^Jh|\leqslant\tilde{C}<1$ for all multi-indices $|J|\leqslant|I|$ and vector fields $Z\in\mathfrak{L}$. Here the ``$J$" in (\ref{48}) is the same as that in (\ref{36}).
\end{pro}
\textbf{Proof.} For simplicity, we only show (\ref{47}). The other estimates can be deduced by the same approach.

Reviewing the definition of $F$ gives
\begin{eqnarray*}
Z^IF_{\mu\nu}&=&Z^I\{P(\partial_{\mu}h,\partial_{\nu}h)+Q_{\mu\nu}(\partial h,\partial h)+G_{\mu\nu}(h)(\partial h,\partial h)\}\\
&&+\mbox{Term}_1+\mbox{Term}_2+\mbox{Term}_3+\mbox{Term}_4+\mbox{Term}_5+\mbox{Term}_6+\mbox{Term}_7,
\end{eqnarray*}
where
\begin{eqnarray*}
\mbox{Term}_1:=2Z^I\{(h^{\alpha\beta}-m^{\alpha\beta})\langle\partial_{\alpha}A_{\mu}-\partial_{\mu}A_{\alpha}+[A_{\alpha},A_{\mu}],\partial_{\beta}A_{\nu}-\partial_{\nu}A_{\beta}+[A_{\beta},A_{\nu}]\rangle\},
\end{eqnarray*}
\begin{eqnarray*}
\mbox{Term}_2:=-2Z^I\{O^{\alpha\beta}(h^2)\langle\partial_{\alpha}A_{\mu}-\partial_{\mu}A_{\alpha}+[A_{\alpha},A_{\mu}],\partial_{\beta}A_{\nu}-\partial_{\nu}A_{\beta}+[A_{\beta},A_{\nu}]\rangle\},
\end{eqnarray*}
\begin{eqnarray*}
\mbox{Term}_3&:=&\frac{1}{2}Z^I\big\{(m^{\alpha\rho}m^{\beta\sigma}h_{\mu\nu}+m^{\alpha\rho}m^{\beta\sigma}m_{\mu\nu}-h^{\alpha\rho}m^{\beta\sigma}m_{\mu\nu}-m^{\alpha\rho}h^{\beta\sigma}m_{\mu\nu})\\
&&\times\langle\partial_{\alpha}A_{\beta}-\partial_{\beta}A_{\alpha}+[A_{\alpha},A_{\beta}],\partial_{\rho}A_{\sigma}-\partial_{\sigma}A_{\rho}+[A_{\rho},A_{\sigma}]\rangle\big\},\nonumber
\end{eqnarray*}
\begin{eqnarray*}
\mbox{Term}_4:=Z^I\left\{O^{\alpha\rho\beta\sigma}_{\mu\nu}(h^2)\langle\partial_{\alpha}A_{\beta}-\partial_{\beta}A_{\alpha}+[A_{\alpha},A_{\beta}],\partial_{\rho}A_{\sigma}-\partial_{\sigma}A_{\rho}+[A_{\rho},A_{\sigma}]\rangle\right\},
\end{eqnarray*}
\begin{eqnarray*}
\mbox{Term}_5:=-Z^I\{(h_{\mu\nu}+m_{\mu\nu})V(|\Phi|^2)\},
\end{eqnarray*}
\begin{eqnarray*}
\mbox{Term}_6:=-Z^I\{(\partial_{\mu}\Phi-\sqrt{-1}A^{I_1}_{\mu}\sigma_{I_1}\Phi/2)^{\dag}(\partial_{\nu}\Phi-\sqrt{-1}A^{I_2}_{\nu}\sigma_{I_2}\Phi/2)\}
\end{eqnarray*}
and
\begin{eqnarray*}
\mbox{Term}_7:=-Z^I\{(\partial_{\nu}\Phi-\sqrt{-1}A^{I_2}_{\nu}\sigma_{I_2}\Phi/2)^{\dag}(\partial_{\mu}\Phi-\sqrt{-1}A^{I_1}_{\mu}\sigma_{I_1}\Phi/2)\}.
\end{eqnarray*}
(9.28) of \cite{LR2010} tells us that, if $|Z^Jh|\leqslant\tilde{C}<1$ for all multi-indices $|J|\leqslant|I|$ and vector fields $Z\in\mathfrak{L}$, one will obtain
\begin{eqnarray}\label{45}
|Z^I(P+Q+G)|&\lesssim&\sum\limits_{|J|+|K|\leqslant|I|}(|\partial Z^Jh|_{\mathcal{T}\mathcal{U}}\cdot|\partial Z^Kh|_{\mathcal{T}\mathcal{U}}+|\bar{\partial}Z^Jh|\cdot|\partial Z^Kh|)\nonumber\\
&&+\sum\limits_{|J|+|K|\leqslant|I|-1}|\partial Z^Jh|_{\mathcal{L}\mathcal{T}}\cdot|\partial Z^Kh|+\sum\limits_{|J|+|K|\leqslant|I|-2}|\partial Z^Jh|\cdot|\partial Z^Kh|\\
&&+\sum\limits_{|J_1|+|J_2|+|J_3|\leqslant|I|}|Z^{J_3}h|\cdot|\partial Z^{J_2}h|\cdot|\partial Z^{J_1}h|.\nonumber
\end{eqnarray}
Moreover, it is easy to get
\begin{eqnarray*}
|\mbox{Term}_1|\lesssim\sum\limits_{|J_1|+|J_2|\leqslant|I|}\Bigg|Z^{J_1}\partial A+\sum\limits_{|L_1|+|L_2|=|J_1|}|Z^{L_1}A|\cdot|Z^{L_2}A|\Bigg|\cdot\Bigg|Z^{J_2}\partial A+\sum\limits_{|L_3|+|L_4|=|J_2|}|Z^{L_3}A|\cdot|Z^{L_4}A|\Bigg|.
\end{eqnarray*}
From induction argument it follows that for any multi-index $I$, there exist a set of universal constants $\{C_J:0\leqslant|J|\leqslant|I|\}$ such that
\begin{eqnarray}\label{46}
Z^I\partial_{\alpha}=\sum\limits_{0\leqslant|J|\leqslant|I|}C_J\cdot\partial_{\alpha}Z^J,
\end{eqnarray}
implying
\begin{eqnarray*}
|\mbox{Term}_1|&\lesssim&\sum\limits_{|J_1|+|J_2|\leqslant|I|}|\partial Z^{J_1}A|\cdot|\partial Z^{J_2}A|+\sum\limits_{|J_1|+|J_2|+|J_3|\leqslant|I|}|\partial Z^{J_1}A|\cdot|Z^{J_2}A|\cdot|Z^{J_3}A|\\
&&+\sum\limits_{|J_1|+|J_2|+|J_3|+|J_4|\leqslant|I|}|Z^{J_1}A|\cdot|Z^{J_2}A|\cdot|Z^{J_3}A|\cdot|Z^{J_4}A|.
\end{eqnarray*}
The same method leads to
\begin{eqnarray*}
|\mbox{Term}_2|&\lesssim&\sum\limits_{|J_1|+|J_2|\leqslant|I|}|\partial Z^{J_1}A|\cdot|\partial Z^{J_2}A|+\sum\limits_{|J_1|+|J_2|+|J_3|\leqslant|I|}|\partial Z^{J_1}A|\cdot|Z^{J_2}A|\cdot|Z^{J_3}A|\\
&&+\sum\limits_{|J_1|+|J_2|+|J_3|+|J_4|\leqslant|I|}|Z^{J_1}A|\cdot|Z^{J_2}A|\cdot|Z^{J_3}A|\cdot|Z^{J_4}A|,
\end{eqnarray*}
\begin{eqnarray*}
|\mbox{Term}_3|&\lesssim&\sum\limits_{|J_1|+|J_2|\leqslant|I|}|\partial Z^{J_1}A|\cdot|\partial Z^{J_2}A|+\sum\limits_{|J_1|+|J_2|+|J_3|\leqslant|I|}|\partial Z^{J_1}A|\cdot|Z^{J_2}A|\cdot|Z^{J_3}A|\\
&&+\sum\limits_{|J_1|+|J_2|+|J_3|+|J_4|\leqslant|I|}|Z^{J_1}A|\cdot|Z^{J_2}A|\cdot|Z^{J_3}A|\cdot|Z^{J_4}A|
\end{eqnarray*}
and
\begin{eqnarray*}
|\mbox{Term}_4|&\lesssim&\sum\limits_{|J_1|+|J_2|\leqslant|I|}|\partial Z^{J_1}A|\cdot|\partial Z^{J_2}A|+\sum\limits_{|J_1|+|J_2|+|J_3|\leqslant|I|}|\partial Z^{J_1}A|\cdot|Z^{J_2}A|\cdot|Z^{J_3}A|\\
&&+\sum\limits_{|J_1|+|J_2|+|J_3|+|J_4|\leqslant|I|}|Z^{J_1}A|\cdot|Z^{J_2}A|\cdot|Z^{J_3}A|\cdot|Z^{J_4}A|.
\end{eqnarray*}
On the other hand, by elementary computation we obtain
\begin{eqnarray*}
|\mbox{Term}_5|&\lesssim&\sum\limits_{|J_1|+|I_1|+\cdots+|I_{2s}|=|I|}|Z^{J_1}h|\cdot|V^{(s)}(|\Phi|^2)|\cdot\prod\limits_{i=1}^{2s}|Z^{I_i}\Phi|\\
&&+\sum\limits_{|I_1|+\cdots+|I_{2s}|=|I|}|V^{(s)}(|\Phi|^2)|\cdot\prod\limits_{i=1}^{2s}|Z^{I_i}\Phi|.
\end{eqnarray*}
Furthermore, the following estimates are not difficult to get:
\begin{eqnarray*}
|\mbox{Term}_6|&\lesssim&\sum\limits_{|J_1|+|J_2|\leqslant|I|}|\partial Z^{J_1}\Phi|\cdot|\partial Z^{J_2}\Phi|+\sum\limits_{|J_1|+|J_2|+|J_3|\leqslant|I|}|Z^{J_1}A|\cdot|Z^{J_2}\Phi|\cdot|\partial Z^{J_3}\Phi|\\
&&+\sum\limits_{|J_1|+|J_2|+|J_3|+|J_4|=|I|}|Z^{J_1}A|\cdot|Z^{J_2}A|\cdot|Z^{J_3}\Phi|\cdot|Z^{J_4}\Phi|
\end{eqnarray*}
and
\begin{eqnarray*}
|\mbox{Term}_7|&\lesssim&\sum\limits_{|J_1|+|J_2|\leqslant|I|}|\partial Z^{J_1}\Phi|\cdot|\partial Z^{J_2}\Phi|+\sum\limits_{|J_1|+|J_2|+|J_3|\leqslant|I|}|Z^{J_1}A|\cdot|Z^{J_2}\Phi|\cdot|\partial Z^{J_3}\Phi|\\
&&+\sum\limits_{|J_1|+|J_2|+|J_3|+|J_4|=|I|}|Z^{J_1}A|\cdot|Z^{J_2}A|\cdot|Z^{J_3}\Phi|\cdot|Z^{J_4}\Phi|.
\end{eqnarray*}
Then the result of this proposition follows.
\endproof

\section{Energy estimates for the EYMH equations}\label{section6}
In this section we prove the following result.

\begin{thm}\label{thm8.1}
Let $h_{\mu\nu}=g_{\mu\nu}-m_{\mu\nu}$, $A_{\beta}$ and $\Phi$ be a local in time solution to (\ref{23}), (\ref{36}) and $(\ref{99})$ respectively satisfying the wave coordinates and Lorentzian gauge conditions on the interval $[0,T^*)$. Suppose also $\gamma\in(0,1/2)$ and $\mu\in(0,1/2)$. Assume that we have the following estimates for $t\in[0,T^*)$ and all multi-indices $|I|\leqslant N-4$:
\begin{eqnarray}\label{75}
|\partial H|_{\mathcal{T}\mathcal{U}}+\frac{|H|_{\mathcal{L}\mathcal{T}}}{1+|q|}+\frac{|ZH|_{\mathcal{L}\mathcal{L}}}{1+|q|}\lesssim\varepsilon,
\end{eqnarray}
\begin{eqnarray}\label{76}
|\partial Z^Ih|+\frac{|Z^Ih|}{1+|q|}+|\bar{\partial}Z^Ih|\leqslant\left\{
\begin{array}{rcl}
&C\varepsilon(1+|q|)^{-2-\gamma}        &\quad\text{when $q>0$,}\\
&C\varepsilon(1+|q|)^{-3/2}         &\quad\text{when $q<0$,}
\end{array} \right.
\end{eqnarray}
\begin{eqnarray}\label{78}
|\partial Z^IA|+\frac{|Z^IA|}{1+|q|}+|\bar{\partial}Z^IA|\leqslant\left\{
\begin{array}{rcl}
&C\varepsilon(1+|q|)^{-2-\gamma}        &\quad\text{when $q>0$,}\\
&C\varepsilon(1+|q|)^{-3/2}         &\quad\text{when $q<0$,}
\end{array} \right.
\end{eqnarray}
\begin{eqnarray}\label{103}
|\partial Z^I\Phi|+\frac{|Z^I\Phi|}{1+|q|}+|\bar{\partial}Z^I\Phi|\leqslant\left\{
\begin{array}{rcl}
&C\varepsilon(1+|q|)^{-2-\gamma}        &\quad\text{when $q>0$,}\\
&C\varepsilon(1+|q|)^{-3/2}         &\quad\text{when $q<0$,}
\end{array} \right.
\end{eqnarray}
and
\begin{eqnarray}\label{77}
\mathcal{E}_N(0)\leqslant\varepsilon.
\end{eqnarray}
Then there is a positive constant $C'$ dependent of $T$ such that we have the energy estimate
\begin{eqnarray*}
\mathcal{E}_N(t)\leqslant C'\varepsilon^2,
\end{eqnarray*}
for all $t\in[0,T^*)$.
\end{thm}

Assuming the conclusions of Theorem \ref{thm8.1} for a moment we finish the proof of the main Theorem \ref{thm6.1}.
\subsection{End of the proof of Theorem \ref{thm6.1}}
Recall that $T^*$ was defined as the maximal time with the property that the bound
$$\mathcal{E}_N(t)\leqslant2C\varepsilon$$
holds for all $t\in[0,T^*)$. Direct check shows that the estimates of Theorem \ref{thm6.2} imply the assumption (\ref{75})-(\ref{103}). The conclusion of Theorem \ref{thm8.1} states that the energy
$$\mathcal{E}_N(t)\leqslant C'\varepsilon^2,\s\s\s\s\forall t\in[0,T^*).$$
Thus choosing a sufficiently small $\varepsilon>0$ we can show that $\mathcal{E}_N(t)\leqslant C\varepsilon$ thus contracting the maximality of $T^*$ and consequently proving that $(g,A,\Phi)$ is a global solution. Therefore, it remains to prove Theorem \ref{thm8.1}.

\subsection{Proof of Theorem \ref{thm8.1}}
Recall that $h_{\mu\nu}$, $A_{\beta}$ and $\Phi$ satisfy the wave equations $\overset{\sim}{\Box}_gh_{\mu\nu}=F_{\mu\nu}$,     $\overset{\sim}{\Box}_gA_{\beta}=J_{\beta}$ and $\overset{\sim}{\Box}_g\Phi=W$ respectively. Our goal is to compute the energy norms of $Z^Ih$, $Z^IA$ and $Z^I\Phi$, where $Z\in\mathfrak{L}$.

From (11.10) of \cite{LR2010} it follows that
$$
\overset{\sim}{\Box}_gZ^Ih_{\mu\nu}=F^{I}_{\mu\nu}
$$
with
$$
F^{I}:=\hat{Z}^IF-D^I,\s\s\s\s D^I:=\hat{Z}^I\overset{\sim}{\Box}_gh^1-\overset{\sim}{\Box}_gZ^Ih^1,\s\s\mbox{and}\s\s \hat{Z}:=Z+c_Z.
$$
Similarly, we also have
$$\overset{\sim}{\Box}_gZ^IA=\hat{Z}^IJ-W^I\s\s\s\s\mbox{and}\s\s\s\s\overset{\sim}{\Box}_gZ^I\Phi=\hat{Z}^IW-L^I,$$
where $W^I:=\hat{Z}^I\overset{\sim}{\Box}_gA-\overset{\sim}{\Box}_gZ^IA$ and $L^I:=\hat{Z}^I\overset{\sim}{\Box}_g\Phi-\overset{\sim}{\Box}_gZ^I\Phi$. (11.13) of \cite{LR2010} tells us
\begin{eqnarray}\label{80}
&&\int_{\Sigma_t}|\partial Z^Ih|^2w+\int_0^t\int_{\Sigma_{\tau}}|\bar{\partial}Z^Ih|^2w'\\
&\lesssim&\int_{\Sigma_0}|\partial Z^I h|^2w+\int_0^t\int_{\Sigma_{\tau}}\Big\{\frac{\varepsilon|\partial Z^Ih|^2}{1+\tau}w+\frac{w\cdot(1+\tau)}{\varepsilon}(|\hat{Z}^IF|^2+|D^I|^2)\Big\},\nonumber\\
&\lesssim&\int_{\Sigma_0}|\partial Z^I h|^2w+\int_0^t\int_{\Sigma_{\tau}}\Big\{\varepsilon|\partial Z^Ih|^2w+\frac{w}{\varepsilon}(|\hat{Z}^IF|^2+|D^I|^2)\Big\},\nonumber
\end{eqnarray}
where $\Sigma_t:=\mathbb{R}^3\times\{t\}$. Applying the same methods yields
\begin{eqnarray}\label{81}
&&\int_{\Sigma_t}|\partial Z^IA|^2w+\int_0^t\int_{\Sigma_{\tau}}|\bar{\partial}Z^IA|^2w'\\
&\lesssim&\int_{\Sigma_0}|\partial Z^IA|^2w+\int_0^t\int_{\Sigma_{\tau}}\left\{\varepsilon|\partial Z^IA|^2w+\frac{w}{\varepsilon}(|\hat{Z}^IJ|^2+|W^I|^2)\right\}.\nonumber
\end{eqnarray}
and
\begin{eqnarray}\label{104}
&&\int_{\Sigma_t}|\partial Z^I\Phi|^2w+\int_0^t\int_{\Sigma_{\tau}}|\bar{\partial}Z^I\Phi|^2w'\\
&\lesssim&\int_{\Sigma_0}|\partial Z^I\Phi|^2w+\int_0^t\int_{\Sigma_{\tau}}\left\{\varepsilon|\partial Z^I\Phi|^2w+\frac{w}{\varepsilon}(|\hat{Z}^IW|^2+|L^I|^2)\right\}.\nonumber
\end{eqnarray}
We begin with the following estimates on the inhomogeneous terms $F$, $J$ and $W$.
\begin{lem}\label{lem8.2}
Under the assumptions of Theorem \ref{thm8.1}, we have
\begin{eqnarray}\label{82}
|Z^IF|&\lesssim&\sum\limits_{|K|\leqslant|I|}\left\{\varepsilon|\partial Z^Kh|+\varepsilon(1+|q|)^{-3/2}|\bar{\partial}Z^Kh|+\varepsilon^2\frac{|Z^Kh|}{(1+|q|)^3}\right\}\nonumber\\
&&+\varepsilon(1+|q|)^{-1}\sum\limits_{|K|\leqslant|I|}(|\partial Z^KA|+|\partial Z^K\Phi|)+\varepsilon^{2\tilde{\gamma}}(1+|q|)^{-\tilde{\gamma}}\\
&&+\varepsilon^3(1+|q|)^{-3/2}\sum\limits_{|K|\leqslant|I|}(|Z^KA|+|Z^K\Phi|+|Z^Kh|)\nonumber
\end{eqnarray}
\begin{eqnarray}\label{86}
|Z^IJ|&\lesssim&\sum\limits_{|K|\leqslant|I|}|\partial Z^Kh|\cdot\varepsilon(1+|q|)^{-1}+\varepsilon^2(1+|q|)^{-1}\sum\limits_{|K|\leqslant|I|}|Z^KA|\nonumber\\
&&+\varepsilon\sum\limits_{|K|\leqslant|I|}(|\partial Z^KA|+|\partial Z^K\Phi|)(1+|q|)^{-1/2}
\end{eqnarray}
and
\begin{eqnarray}\label{105}
|Z^IW|&\lesssim&\sum\limits_{|K|\leqslant|I|}|\partial Z^K\Phi|\cdot\varepsilon(1+|q|)^{-1/2}+\varepsilon^2(1+|q|)^{-1}\sum\limits_{|K|\leqslant|I|}|Z^K\Phi|
\end{eqnarray}
\end{lem}

\textbf{Proof.} We only prove (\ref{82}), since the other cases are easy(Note that throughout the process we have to use the assumption $\tilde{\gamma}>3/2+\gamma$). According to Proposition \ref{pro6.3} we have
\begin{eqnarray*}
|Z^IF|&\lesssim&\mbox{Term}+\sum\limits_{|J_1|+|J_2|\leqslant|I|}|\partial Z^{J_1}A|\cdot|\partial Z^{J_2}A|+\sum\limits_{|J_1|+|J_2|+|J_3|\leqslant|I|}|\partial Z^{J_1}A|\cdot|Z^{J_2}A|\cdot|Z^{J_3}A|\\
&&+\sum\limits_{|J_1|+|J_2|+|J_3|+|J_4|\leqslant|I|}|Z^{J_1}A|\cdot|Z^{J_2}A|\cdot|Z^{J_3}A|\cdot|Z^{J_4}A|\\
&&+\sum\limits_{|I_1|+\cdots+|I_{2s}|=|I|}|V^{(s)}(|\Phi|^2)|\cdot\prod\limits_{i=1}^{2s}|Z^{I_i}\Phi|\nonumber\\
&&+\sum\limits_{|J_1|+|I_1|+\cdots+|I_{2s}|=|I|}|Z^{J_1}h|\cdot|V^{(s)}(|\Phi|^2)|\cdot\prod\limits_{i=1}^{2s}|Z^{I_i}\Phi|\nonumber\\
&&+\sum\limits_{|J_1|+|J_2|\leqslant|I|}|\partial Z^{J_1}\Phi|\cdot|\partial Z^{J_2}\Phi|+\sum\limits_{|J_1|+|J_2|+|J_3|\leqslant|I|}|Z^{J_1}A|\cdot|Z^{J_2}\Phi|\cdot|\partial Z^{J_3}\Phi|\nonumber\\
&&+\sum\limits_{|J_1|+|J_2|+|J_3|+|J_4|=|I|}|Z^{J_1}A|\cdot|Z^{J_2}A|\cdot|Z^{J_3}\Phi|\cdot|Z^{J_4}\Phi|,\nonumber
\end{eqnarray*}
where
\begin{eqnarray*}
\mbox{Term}&:=&\sum\limits_{|J|+|K|\leqslant|I|}(|\partial Z^Jh|_{\mathcal{T}\mathcal{U}}\cdot|\partial Z^Kh|_{\mathcal{T}\mathcal{U}}+|\bar{\partial}Z^Jh|\cdot|\partial Z^Kh|)\nonumber\\
&&+\sum\limits_{|J|+|K|\leqslant|I|-1}|\partial Z^Jh|_{\mathcal{L}\mathcal{T}}\cdot|\partial Z^Kh|+\sum\limits_{|J|+|K|\leqslant|I|-2}|\partial Z^Jh|\cdot|\partial Z^Kh|\\
&&+\sum\limits_{|J_1|+|J_2|+|J_3|\leqslant|I|}|Z^{J_3}h|\cdot|\partial Z^{J_2}h|\cdot|\partial Z^{J_1}h|.
\end{eqnarray*}
From Theorem \ref{thm6.2} it follows that
\begin{eqnarray}\label{84}
\mbox{Term}&\lesssim&\sum\limits_{|K|\leqslant|I|}\left\{\varepsilon|\partial Z^Kh|+\varepsilon(1+|q|)^{-3/2}|\bar{\partial}Z^Kh|+\varepsilon^2\frac{|Z^Kh|}{(1+|q|)^3}\right\}.
\end{eqnarray}
In addition, (\ref{78}) implies
\begin{eqnarray}\label{83}
\sum\limits_{|I_1|+|I_2|\leqslant|I|}|\partial Z^{I_1}A|\cdot|\partial Z^{I_2}A|\lesssim\varepsilon(1+|q|)^{-3/2}\sum\limits_{|K|\leqslant|I|}|\partial Z^KA|.
\end{eqnarray}
Furthermore, from (\ref{78}) we get
\begin{eqnarray*}
|Z^IA|\leqslant C\varepsilon(1+|q|)^{-1/2}.
\end{eqnarray*}
Hence, one can obtain
\begin{eqnarray*}
\sum\limits_{|J_1|+|J_2|+|J_3|\leqslant|I|}|\partial Z^{J_1}A|\cdot|Z^{J_2}A|\cdot|Z^{J_3}A|\leqslant C^2\varepsilon^2(1+|q|)^{-1}\sum\limits_{|K|\leqslant|I|}|\partial Z^KA|.
\end{eqnarray*}
The same method yields the following inequalities:
\begin{eqnarray*}
\sum\limits_{|J_1|+|J_2|+|J_3|+|J_4|\leqslant|I|}|Z^{J_1}A|\cdot|Z^{J_2}A|\cdot|Z^{J_3}A|\cdot|Z^{J_4}A|\leqslant C^3\varepsilon^3(1+|q|)^{-3/2}\sum\limits_{|K|\leqslant|I|}|Z^KA|,
\end{eqnarray*}
\begin{eqnarray*}
\sum\limits_{|I_1|+\cdots+|I_{2s}|=|I|}|V^{(s)}(|\Phi|^2)|\cdot\prod\limits_{i=1}^{2s}|Z^{I_i}\Phi|\leqslant C^{2\tilde{\gamma}}\varepsilon^{2\tilde{\gamma}}(1+|q|)^{-\tilde{\gamma}},
\end{eqnarray*}
\begin{eqnarray*}
\sum\limits_{|J_1|+|I_1|+\cdots+|I_{2s}|=|I|}|Z^{J_1}h|\cdot|V^{(s)}(|\Phi|^2)|\cdot\prod\limits_{i=1}^{2s}|Z^{I_i}\Phi|\leqslant C^{2\tilde{\gamma}}\varepsilon^{2\tilde{\gamma}}(1+|q|)^{-\tilde{\gamma}}\sum\limits_{|K|\leqslant|I|}|Z^Kh|,
\end{eqnarray*}
\begin{eqnarray*}
\sum\limits_{|J_1|+|J_2|\leqslant|I|}|\partial Z^{J_1}\Phi|\cdot|\partial Z^{J_2}\Phi|\leqslant C\varepsilon(1+|q|)^{-3/2}\sum\limits_{|K|\leqslant|I|}|\partial Z^K\Phi|,
\end{eqnarray*}
\begin{eqnarray*}
\sum\limits_{|J_1|+|J_2|+|J_3|\leqslant|I|}|Z^{J_1}A|\cdot|Z^{J_2}\Phi|\cdot|\partial Z^{J_3}\Phi|\leqslant C^2\varepsilon^2(1+|q|)^{-1}\sum\limits_{|K|\leqslant|I|}|\partial Z^K\Phi|,
\end{eqnarray*}
and
\begin{eqnarray*}
\sum\limits_{|J_1|+|J_2|+|J_3|+|J_4|=|I|}|Z^{J_1}A|\cdot|Z^{J_2}A|\cdot|Z^{J_3}\Phi|\cdot|Z^{J_4}\Phi|\leqslant C^3\varepsilon^3(1+|q|)^{-3/2}\sum\limits_{|K|\leqslant|I|}|Z^K\Phi|.
\end{eqnarray*}
Combining the above estimates gives (\ref{82}).
\endproof

\begin{lem}\label{lem8.3}
Under the assumptions of Theorem \ref{thm8.1}, one can get
\begin{eqnarray}
&&\varepsilon^{-1}\int_0^t\int|Z^IF|^2wdxd\tau\lesssim\varepsilon\sum\limits_{|K|\leqslant|I|}\int_0^t\int(|\partial Z^Kh|^2+|\partial Z^KA|^2+|\partial Z^K\Phi|^2)wdxd\tau\\
&&+\varepsilon^{4\tilde{\gamma}-1}+\sum\limits_{|K|\leqslant|I|}\varepsilon\int_0^t\int|\bar{\partial} Z^Kh|^2w'dxd\tau,\nonumber
\end{eqnarray}
\begin{equation}
\varepsilon^{-1}\int_0^t\int|Z^IJ|^2wdxd\tau\lesssim\sum\limits_{|K|\leqslant|I|}\varepsilon\int_0^t\int(|\partial Z^Kh|^2+|\partial Z^KA|^2+|\partial Z^K\Phi|^2)wdxd\tau
\end{equation}
and
\begin{eqnarray}
\varepsilon^{-1}\int_0^t\int|Z^IW|^2wdxd\tau\lesssim\varepsilon\cdot\sum\limits_{|K|\leqslant|I|}\int_0^t\int|\partial Z^K\Phi|^2wdxd\tau.
\end{eqnarray}
\end{lem}
\textbf{Proof.} Throughout the process we have to apply Corollary 13.3 of \cite{LR2010}(to transform $Z^Kh$, $Z^KA$ and $Z^K\Phi$ into $\partial Z^Kh$, $\partial Z^KA$ and $\partial Z^K\Phi$ respectively), the inequality $1\leqslant1+\tau\leqslant1+T$(Note that the constant $C$ relies on the maximal existence time $T$) and the assumption $\tilde{\gamma}>3/2+\gamma$(to ensure that some integral is finite). Indeed, if the integral is denoted by $In$, we can give its specific expression
\begin{eqnarray*}
In:=\varepsilon^{4\tilde{\gamma}-1}\int_{\mathbb{R}^3}(1+|q|)^{-2\tilde{\gamma}}w\,dx^1dx^2dx^3.
\end{eqnarray*}
Since $w\lesssim(1+|q|)^{1+2\gamma}$, under polar coordinates system it is easy to get
\begin{eqnarray*}
In\lesssim\varepsilon^{4\tilde{\gamma}-1}\int_{\mathbb{S}^2}d\mathbb{S}^2\int_0^{\infty}(1+|q|)^{1+2\gamma-2\tilde{\gamma}}r^2\,dr.
\end{eqnarray*}
Hence, the assumption $\tilde{\gamma}>3/2+\gamma$ implies $In<\infty$.

The other part of the proof is similar to that of Lemma 11.3 of \cite{LR2010}. So we omit it.
\endproof

Now we deal with $D^I$, $W^I$ and $L^I$.

\begin{lem}\label{lem8.4}
Under the assumptions of Theorem \ref{thm8.1}, we have
\begin{equation}\label{88}
\varepsilon^{-1}\int_0^t\int|D^I|^2wdxd\tau\lesssim\varepsilon\sum\limits_{|K|\leqslant|I|}\int_0^t\int\left(|\partial Z^Kh|^2w+|\bar{\partial}Z^Kh|^2w'\right)dxd\tau+\varepsilon^3,
\end{equation}
\begin{eqnarray}\label{89}
\varepsilon^{-1}\int_0^t\int|W^I|^2wdxd\tau&\lesssim&\varepsilon\sum\limits_{|K|\leqslant|I|}\int_0^t\int\left(|\partial Z^Kh|^2w+|\bar{\partial}Z^Kh|^2w'\right)dxd\tau\\
&&+\varepsilon^3+\varepsilon\sum\limits_{|K|\leqslant|I|}\int_0^t\int|\partial Z^KA|^2w\,dxd\tau\nonumber
\end{eqnarray}
and
\begin{eqnarray}\label{106}
\varepsilon^{-1}\int_0^t\int|L^I|^2wdxd\tau&\lesssim&\varepsilon\sum\limits_{|K|\leqslant|I|}\int_0^t\int\left(|\partial Z^Kh|^2w+|\bar{\partial}Z^Kh|^2w'\right)dxd\tau\\
&&+\varepsilon^3+\varepsilon\sum\limits_{|K|\leqslant|I|}\int_0^t\int|\partial Z^K\Phi|^2w\,dxd\tau\nonumber
\end{eqnarray}
\end{lem}
\textbf{Proof.} We only prove (\ref{106}) since the other estimates follow from the same methods. According to Proposition 5.3 of \cite{LR2010} we arrive at
\begin{eqnarray}\label{90}
&&|\overset{\sim}{\Box}_gZ^I\Phi-\hat{Z}^I\overset{\sim}{\Box}_g\Phi|\\
&\lesssim&\sum\limits_{|K|\leqslant|I|}\sum\limits_{|J|+(|K|-1)_+\leqslant|I|}\left(\frac{|Z^JH|}{1+|q|}+\frac{|Z^JH|_{\mathcal{L}\mathcal{L}}}{1+|q|}\right)|\partial Z^K\Phi|\nonumber\\
&&+\sum\limits_{|K|\leqslant|I|}\left(\sum\limits_{|J|+(|K|-1)_+\leqslant|I|-1}\frac{|Z^JH|_{\mathcal{L}\mathcal{T}}}{1+|q|}+\sum\limits_{|J|+(|K|-1)_+\leqslant|I|-2}\frac{|Z^JH|}{1+|q|}\right)|\partial Z^K\Phi|.\nonumber
\end{eqnarray}
Our goal is to obtain the estimate for the quantity
$$
\sum\limits_{|I|\leqslant N}\int_0^t\int|\overset{\sim}{\Box}_gZ^I\Phi-\hat{Z}^I\overset{\sim}{\Box}_g\Phi|^2w\,dxd\tau.
$$
Let us first deal with the terms in (\ref{90}) with $|K|\leqslant N-4$. In this case we use the decay estimate (\ref{103}). It is clear that now we only have to consider the expression
\begin{equation}\label{91}
\begin{aligned}
&\sum\limits_{\substack{|J|\leqslant|I|\\|J'|\leqslant|I|-1\\|K|\leqslant|I|-2}}\int_0^t\int\left\{\frac{|Z^JH|^2}{(1+|q|)^2}+\frac{|Z^JH|^2_{\mathcal{L}\mathcal{L}}+|Z^{J'}H|^2_{\mathcal{L}\mathcal{T}}+|Z^KH|^2}{(1+|q|)^2}\right\}\varepsilon^2(1+|q|)^{-3}w\,dxd\tau\\
=&\sum\limits_{\substack{|J|\leqslant|I|\\|J'|\leqslant|I|-1\\|K|\leqslant|I|-2}}\int_0^t\int\left\{|Z^JH|^2+|Z^JH|^2_{\mathcal{L}\mathcal{L}}+|Z^{J'}H|^2_{\mathcal{L}\mathcal{T}}+|Z^KH|^2\right\}\varepsilon^2(1+|q|)^{-1}w\,dxd\tau.
\end{aligned}
\end{equation}
From the proof of Lemma 11.5 in \cite{LR2010} it follows that (\ref{91}) is bounded by
\begin{eqnarray*}
&&C\varepsilon^2\sum\limits_{|K|\leqslant|I|}\int_0^t\int\left(\frac{|\partial Z^Kh|^2}{1+\tau}w+|\bar{\partial}Z^Kh|^2w'\right)dxd\tau\\
&&+C\varepsilon^2\sum\limits_{|K|\leqslant|I|-1}\int_0^t\int\frac{|\partial Z^Kh|^2}{(1+\tau)^{1-2C\varepsilon}}wdxd\tau+C\varepsilon^4
\end{eqnarray*}
which is equivalent to
\begin{eqnarray*}
C\varepsilon^2\sum\limits_{|K|\leqslant|I|}\int_0^t\int\left(|\partial Z^Kh|^2w+|\bar{\partial}Z^Kh|^2w'\right)dxd\tau+C\varepsilon^4,
\end{eqnarray*}
where we let the parameter $M$ in the expression $H_0^{\mu\nu}:=-\chi(r/t)\chi(r)M\delta^{\mu\nu}/r$ equal to 0(the expression can be found at the beginning of the proof of Lemma 11.5 in \cite{LR2010}). For more details we refer to the last inequality on page 1460 of \cite{LR2010}.

Returning to (\ref{90}) we now deal with the case $|K|\geqslant N-3$, which implies $|J|\leqslant 4$. By the same way of proving Lemma 11.5 of \cite{LR2010} we arrive at that the contribution of the terms with $|K|\geqslant N-3$ to $|\overset{\sim}{\Box}_gZ^I\Phi-\hat{Z}^I\overset{\sim}{\Box}_g\Phi|$ can be bounded by
$$
\varepsilon\sum\limits_{|K|=|I|}\frac{|\partial Z^K\Phi|}{1+\tau}+\varepsilon\sum\limits_{|K|<|I|}\frac{|\partial Z^K\Phi|}{(1+\tau)^{1-C\varepsilon}},
$$
which is equivalent to
$$
\varepsilon\sum\limits_{|K|\leqslant|I|}|\partial Z^K\Phi|.
$$
\endproof

Now let us finish the proof of Theorem \ref{thm8.1}. Applying (\ref{80}), (\ref{81}) and (\ref{104}) together with Lemma \ref{lem8.3} and Lemma \ref{lem8.4} yields
\begin{equation}\label{93}
\begin{aligned}
&\int_{\Sigma_t}(|\partial Z^Ih|^2+|\partial Z^IA|^2+|\partial Z^I\Phi|^2)w+\int_0^t\int_{\Sigma_{\tau}}(|\bar{\partial} Z^Ih|^2+|\bar{\partial} Z^IA|^2+|\bar{\partial} Z^I\Phi|^2)w'\\
\lesssim&\varepsilon\sum\limits_{|K|\leqslant|I|}\int_0^t\int(|\partial Z^Kh|^2+|\partial Z^KA|^2+|\partial Z^K\Phi|^2)w+\int_{\Sigma_0}(|\partial Z^Ih|^2+|\partial Z^IA|^2+|\partial Z^I\Phi|^2)w\\
&+\varepsilon\sum\limits_{|K|\leqslant|I|}\int_0^t\int(|\bar{\partial} Z^Kh|^2+|\bar{\partial} Z^KA|^2+|\bar{\partial} Z^K\Phi|^2)w'+\varepsilon^3
\end{aligned}
\end{equation}
Denote
$$\tilde{\mathcal{E}}_k(t):=\sup\limits_{0\leqslant\tau\leqslant t}\sum\limits_{\substack{Z\in\mathfrak{L}\\|I|\leqslant k}}\int_{\Sigma_{\tau}}(|\partial Z^Ih|^2+|\partial Z^IA|^2+|\partial Z^I\Phi|^2)w\,dx$$
and
$$
\mathcal{S}_k(t):=\sum\limits_{\substack{Z\in\mathfrak{L}\\|I|\leqslant k}}\int_0^t\int_{\Sigma_{\tau}}(|\bar{\partial} Z^Ih|^2+|\bar{\partial} Z^IA|^2++|\bar{\partial} Z^I\Phi|^2)w'\,dx.
$$
Then we get
\begin{eqnarray}\label{94}
\tilde{\mathcal{E}}_k(t)+\mathcal{S}_k(t)&\lesssim&\tilde{\mathcal{E}}_k(0)+\varepsilon\int_0^t\tilde{\mathcal{E}}_{k}(\tau)d\tau+\varepsilon\mathcal{S}_k(t)+\varepsilon^3,
\end{eqnarray}
which implies
\begin{eqnarray*}
\tilde{\mathcal{E}}_k(t)\leqslant C(T)\varepsilon^2.
\end{eqnarray*}
\endproof

{}

\vspace{1.0cm}

Zonglin Jia

{\small\it Institute of Applied Physics and Computational Mathematics, China Academy of Engineering Physics, Beijing, 100088, P. R. China}

{\small\it Email: 756693084@qq.com}

Boling Guo

{\small\it Institute of Applied Physics and Computational Mathematics, China Academy of Engineering Physics, Beijing, 100088, P. R. China}

{\small\it Email: gbl@iapcm.ac.cn}\\

\end{document}